\documentclass[a4paper]{article}

\usepackage{a4wide}
\usepackage[UKenglish]{babel}
\usepackage{graphicx,subfigure}
	\graphicspath{{./}{figures/}}
\usepackage[colorlinks=true,linkcolor=blue,citecolor=red]{hyperref}
\usepackage{xcolor}
\usepackage{amssymb,amsthm,empheq,bbold}
\usepackage[inline]{enumitem}
\usepackage{multirow}
\usepackage{caption}
\theoremstyle{remark}\newtheorem{remark}{Remark}[section]
\theoremstyle{plain}
\newtheorem{proposition}[remark]{Proposition}

\theoremstyle{definition}

\newcommand{\abs}[1]{\left\lvert#1\right\rvert}
\DeclarePairedDelimiter\ave{\langle}{\rangle}
\newcommand{\bI}{\mathbf{I}}
\newcommand{\cI}{\mathcal{I}}

\newcommand{\bm}{\mathbf{m}}
\newcommand{\Prob}{\operatorname{Prob}}
\newcommand{\bP}{\mathbf{P}}
\newcommand{\R}{\mathbb{R}}
\newcommand{\brho}{\boldsymbol{\rho}}
\newcommand{\bzero}{\mathbf{0}}

\allowdisplaybreaks

\begin{document}
\title{A viral load-based model for epidemic spread on spatial networks}
\author{Nadia Loy\thanks{\texttt{nadia.loy@polito.it}} \and Andrea Tosin\thanks{\texttt{andrea.tosin@polito.it}}}
\date{\small Department of Mathematical Sciences ``G. L. Lagrange'' \\ Politecnico di Torino, Italy}

\maketitle

\begin{abstract}
In this paper, we propose a Boltzmann-type kinetic model of the spreading of an infectious disease on a network. The latter describes the connections among countries, cities or districts depending on the spatial scale of interest. The disease transmission is represented in terms of the viral load of the individuals and is mediated by social contacts among them, taking into account their displacements across the nodes of the network. We formally derive the hydrodynamic equations for the density and the mean viral load of the individuals on the network and we analyse the large-time trends of these quantities with special emphasis on the cases of blow-up or eradication of the infection. By means of numerical tests, we also investigate the impact of confinement measures, such as quarantine or localised lockdown, on the diffusion of the disease on the network.
\end{abstract}
\medskip

\noindent{\bf Keywords:} Boltzmann-type equations, Markov-type jump processes, label switching, commuters, quarantine

\medskip

\noindent{\bf Mathematics Subject Classification:} 35Q20, 35Q70, 35Q84

\section{Introduction}
The spreading of infectious diseases is intimately correlated with patterns of human movement, as clearly demonstrated by empirical observations~\cite{wesolowski2017NC}. Yet, many compartmental epidemiological models, inspired by the celebrated Kermack-McKendrick theory~\cite{kermack1991BMB}, rely on the assumption of homogeneous mixing of the individuals and on the gross number of social contacts among them to provide a big picture of the epidemic trends. Actually, the inclusion of a spatial layer in the mathematical description of epidemic spreading is a quite explored topic in the pertinent literature, cf. e.g.,~\cite{colizza2008JTB,keeling2005JRSI}. Moreover, the recent COVID-19 pandemic has renewed the interest in this issue~\cite{bertaglia2020MMNA,boscheri2020PREPRINT}, given the evident role played by global connections in the worldwide spreading of the epidemic after its localised onset. The spatial layer may be taken into account in essentially two ways. One way consists in introducing a domain, which represents the spatial area of interest, and describing there fluxes of individuals by means of transport-diffusion equations in the spirit of fluid dynamics, cf. e.g.,~\cite{almeida2020PREPRINT,boscheri2020PREPRINT} and~\cite[Chapter~15]{martcheva2015BOOK}. This approach is generally suited to small spatial scales, because, on one hand, it requires to model local movements of the individuals and, on the other hand, it hardly allows one to account for main directions and preferential paths. An alternative way consists in schematising the spatial structure of the problem by means of a \textit{network}, whose nodes represent spatial locations and whose links indicate connections among them. Nodes are regarded as spatially homogeneous, hence within them the epidemiological state of the individuals evolves according to contact-and-transmission rules with no spatial structure. However, since individuals may move from one node to another the description is on the whole spatially inhomogeneous and appears to be particularly suited to model main and preferential paths. In addition to this, such a spatial representation is quite versatile as it may describe equally well both short-range and long-range connections. Contributions in this direction are e.g., the so-called \textit{meta-population models}, cf.~\cite{apolloni2014TBMM,arrigoni2007CHAPTER,barbour2004JMB,colizza2008JTB}. Here, the movements of the individuals are described as ``jumps'' from one node of the network to another following the graph edges and the weights prescribed to them, which are interpreted as mobility rates, see also the recent survey~\cite{zino2021PREPRINT} and the specific application~\cite{parino2020PREPRINT} to the case of the COVID-19 epidemics. Instead, we remark that in the aforementioned paper~\cite{bertaglia2020MMNA} the dynamical fluxes of individuals along the links of the network are explicitly modelled in the spirit of the models of vehicular traffic on road networks~\cite{garavello2016BOOK,garavello2006BOOK}.

As mentioned at the beginning, another issue in standard compartmental epidemiological models is the general lack of specific disease transmission mechanisms. Contagions are usually modelled only on the basis of the gross numbers of susceptible and infected individuals, assuming that the higher these numbers the higher the probability for a susceptible individual to get in contact with an infected one. This is, for instance, the driving mechanism of the celebrated SIS and SIR models then borrowed by many other more sophisticated compartmental models. Also recent models, partly based on more detailed descriptions of social contacts among the individuals, see e.g.,~\cite{dimarco2020PRE,dimarco2021PREPRINT}, still rely on a SIR-like structure for the description of the contagion dynamics. Nevertheless, it is known that some specific determinants, such as the \textit{viral load} carried by the individuals, are at the basis of the infection transmission among the individuals and of its detection through viral tests~\cite{larremore2021SA}. Measurements of these determinants play also a role in the actuation of confinement measures, such as the quarantine of individuals diagnosed as infected. To the best of our knowledge, epidemiological models describing contagion dynamics by means of a viral load variable are still not common in the literature.

Motivated by the considerations above, in this work we provide as main contribution the design of new mathematical models of contagion dynamics on networks based on viral load transmission. In particular, we propose non-compartmental evolutionary models formalised in terms of the density of individuals in each node of the network and the mean viral load there, which yields a direct measure of the extent of the local contagion. In our models, we follow the idea of the jumps across the nodes, thus we do not describe the fluxes of individuals along the links of the network. To build these models from scratch we adopt a \textit{statistical mechanics} point of view through \textit{Boltzmann-type kinetic equations}. This approach allows us to postulate elementary viral load transmission dynamics within the nodes and migration dynamics across the nodes and to upscale them in a formally rigorous way to get the aggregate macroscopic description. We stress that, in our case, a compartmentalisation of the individuals in each node based on their infection condition is not necessary. Indeed, we may rely on the viral load variable to describe contagion dynamics at the individual scale and on the mean viral load to account for the aggregate epidemiological trends. We also remark that kinetic equations on graphs are a still largely under-explored topic in the broad literature on kinetic models of social interactions, see e.g.,~\cite{burger2020PREPRINT} for a very recent proposal. Therefore, our work provides as a by-product also a contribution in this direction.

In more detail, the paper is organised as follows. In Section~\ref{sect:general_model} we introduce the Boltzmann-type kinetic description of social interactions and contagion spread on networks. After obtaining the Boltzmann-type equations accounting for contagion within the nodes and movements across the nodes, we derive exact macroscopic equations for the density of individuals and the mean viral load in each node. We then analyse the large-time trends of these quantities, studying in particular the conditions under which the infection is globally eradicated or blows up depending on the characteristics of the individual contagion dynamics. Next, we specialise the model to the case of commuters, which requires to manage the jumps of the individuals across the nodes in a dedicated manner. In Section~\ref{sect:quarantine} we extend the previous model by introducing the possibility to put node-wise in isolation (quarantine) individuals diagnosed as infected. For this, we take advantage of the viral load as a descriptor of the microscopic state of the individuals and of the techniques developed in~\cite{loy2021KRM_preprint} for kinetic equations with label switching. In essence, in each node we label the individuals as either non-quarantined or quarantined and we allow for changes of label on the basis of individual viral load levels. Such label switches are described in probability as a result of diagnostic tests, which may or may not detect the infection depending on the viral load carried by the individuals being tested. In Section~\ref{sect:numerics} we propose numerical tests of prototypical case studies, which illustrate the type of predictions obtainable from our new models. These tests are also useful to validate the aggregate macroscopic models resulting formally from the kinetic description. Indeed, they show a proper matching between the aggregate trends obtained from the macroscopic models and the original particle dynamics simulated by a Monte Carlo approach. Finally, in Section~\ref{sect:conclusions} we draw some conclusions and we briefly sketch further possible developments.

\section{Social interactions and contagion spread on networks}
\label{sect:general_model}
\begin{figure}[t]
\centering
\includegraphics[width=.7\textwidth]{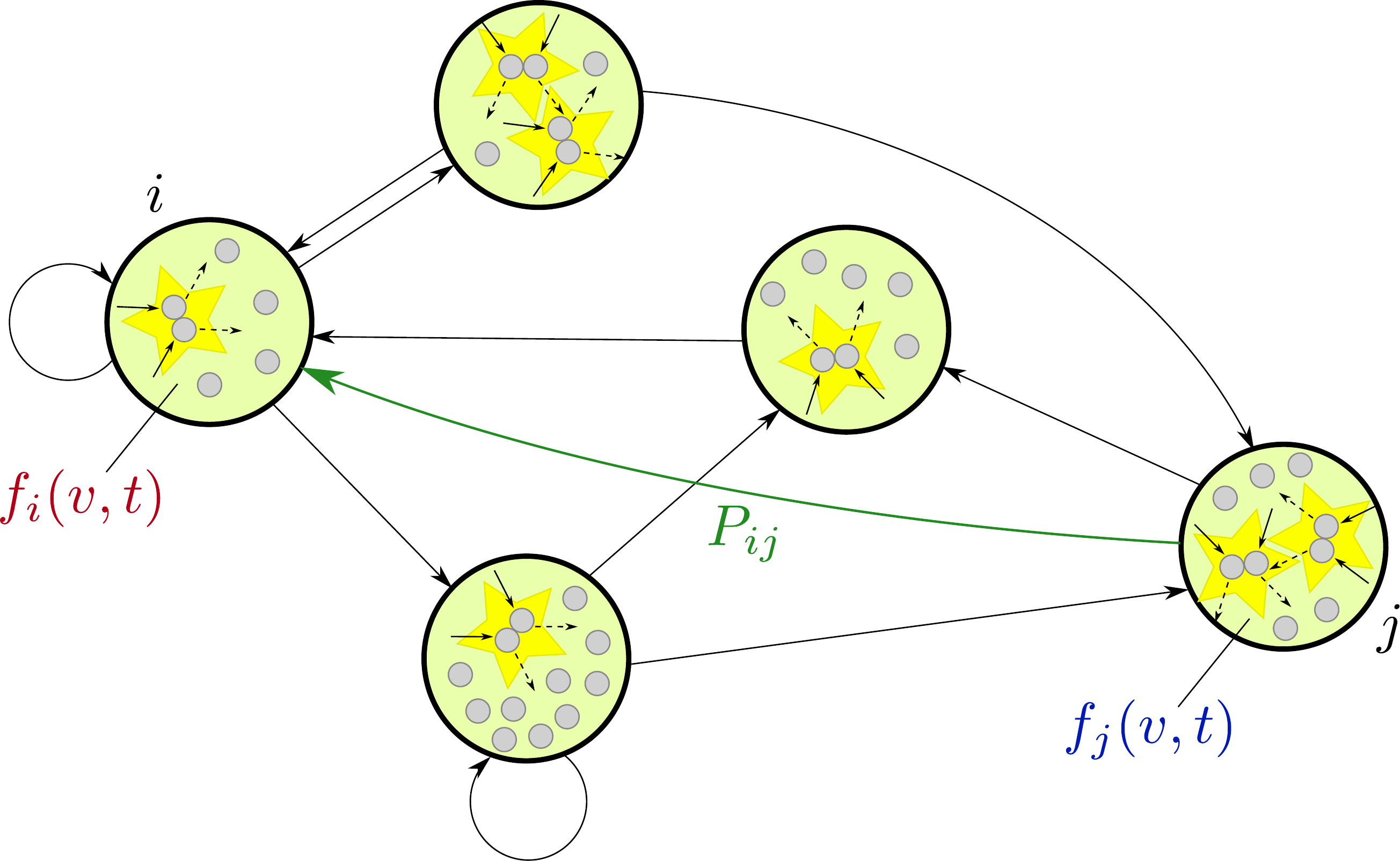}
\caption{A strongly connected graph in which individuals interact within the vertices and may jump from vertex to vertex.}
\label{fig:kingraph}
\end{figure}

\subsection{Boltzmann-type kinetic equations on a graph}
\label{sect:Boltzmann.graph}
Let us consider a large system of interacting individuals migrating on a network (see Figure~\ref{fig:kingraph}), which we model by a graph with a finite number of vertices and edges. The displacements of the individuals from one vertex to another are described as a Markov-type jump process. Labelling each vertex by an index $i\in\cI$, where $\cI$ is a finite ordered index set with $\abs{\cI}=n\in\mathbb{N}$, for instance $\cI=\{1,\,\dots,\,n\}\subset\mathbb{N}$, we may describe such a stochastic jump process by means of a \textit{transition probability}
\begin{equation}
	P_{ij}:=\Prob(j\to i)\in [0,\,1], \qquad i,\,j\in\cI,
	\label{eq:Pij}
\end{equation}
representing the probability to jump from vertex $j$ to vertex $i$. Clearly
\begin{equation}
	\sum_{i\in\cI}P_{ij}=1, \qquad \forall\,j\in\cI,
	\label{eq:P}
\end{equation}
hence the square matrix $\bP:=[P_{ij}]_{i,j\in\cI}\in\R^{n\times n}$, called the \textit{transition matrix}, is left stochastic as its entries are non-negative and its columns sum to one. Since in our application the nodes of the network represent spatial locations, such as districts, cities, countries depending on the scale of interest, it makes sense to assume that the graph is \textit{strongly connected}, i.e. it contains a directed path from every vertex to every other vertex. Equivalently, the matrix $\bP$ is \textit{irreducible}.

Besides their current vertex, individuals are further characterised by a microscopic state $v\in\R_+$, which in the application discussed in this paper represents their \textit{viral load} understood as a measure of their infectivity. The viral load of the individuals evolves in consequence of social contacts, which, in the spirit of the collisional kinetic theory, we schematise as binary interactions of the form
\begin{equation}
	v'=pv+qv_\ast, \qquad v_\ast'=pv_\ast+qv,
	\label{eq:binary_gen_sym}
\end{equation}
where $v,\,v_\ast$ represent the pre-interaction viral loads of two interacting individuals and $v',\,v_\ast'$ their post-interaction viral loads. Moreover, $p,\,q\in\R_+$ are either deterministic or stochastic interaction coefficients. The binary rules~\eqref{eq:binary_gen_sym} express the contagion dynamics, which we assume to be symmetric. Indeed the second rule is obtained from the first one by switching the roles of $v,\,v_\ast$, which means that there is no preferential order of the individuals in an interaction. We also assume that only individuals in the same vertex of the graph may exchange viral load according to~\eqref{eq:binary_gen_sym}, whereas individuals in different vertices do not interact. The reason is that contagion dynamics require a certain proximity to be effective. Notice that~\eqref{eq:binary_gen_sym} assumes implicitly that such dynamics are the same in each vertex. Different contagion dynamics in different vertices might be taken into account by specifying different interaction coefficients $p_i,\,q_i$, $i\in\cI$, from node to node. In particular, we fix
\begin{equation}
	p=1-\nu_1+\eta, \qquad q=\nu_2,
	\label{eq:pq}
\end{equation}
where $\nu_1\in [0,\,1]$ is the physiological decay rate of the viral load of an infected individual, $\eta$ is a random coefficient modelling stochastic fluctuations of the physiological decay rate and $\nu_2\in [0,\,1]$ is the transmission rate of the contagion. Denoting by $\ave{\cdot}$ the expectation with respect to the law of $\eta$, we assume $\ave{\eta}=0$ and moreover $\eta\geq\nu_1-1$ so as to meet the condition $p\in\R_+$, which, together with $q\in\R_+$, guarantees $v',\,v_\ast'\geq 0$ for all $v,\,v_\ast\geq 0$.

From the particle point of view, we model the microscopic dynamics associated with the processes above as follows. We identify a generic individual by its location $X_t\in\cI$ in the network and its viral load $V_t\in\R_+$ at time $t$. Then we describe the evolution of $X_t$, $V_t$ during a time step $\Delta{t}>0$ by means of the following update rules:
\begin{align}
	\begin{aligned}[c]
		X_{t+\Delta{t}} &= (1-\Theta)X_t+\Theta J_t, \\
		V_{t+\Delta{t}} &= \left(1-\Xi\delta_{X_t,X^\ast_t}\right)V_t+\Xi\delta_{X_t,X^\ast_t}V'_t,
	\end{aligned}
	\label{eq:particle.gen}
\end{align}
where:
\begin{enumerate}[label=\roman*)]
\item $\Theta,\,\Xi\in\{0,\,1\}$ are independent Bernoulli random variables discriminating whether a vertex jump or a change in the viral load take place ($\Theta,\,\Xi=1$) or not ($\Theta,\,\Xi=0$) in the time step $\Delta{t}$. In particular, we let
\begin{equation}
	\Prob(\Theta=1)=\chi\Delta{t}, \qquad \Prob(\Xi=1)=\mu\Delta{t},
	\label{eq:Theta.Xi}
\end{equation}
$\chi,\,\mu>0$ being the mobility and social contact rates, and we assume $\Delta{t}\leq\min\{\frac{1}{\chi},\,\frac{1}{\mu}\}$ for consistency;
\item $J_t\in\cI$ is a random variable returning the new vertex after a vertex jump, with
$$ \Prob(J_t=i\vert X_t=j)=P_{ij}, \qquad i,\,j\in\cI; $$
\item $V'_t\in\R_+$ is the new viral load after a social contact. In view of~\eqref{eq:binary_gen_sym} we have $V'_t=pV_t+qV^\ast_t$, where $V^\ast_t$ denotes the viral load at time $t$ of the other individual participating in the interaction;
\item $\delta_{X_t,X^\ast_t}$ is the Kronecker delta:
$$	\delta_{X_t,X^\ast_t}:=
		\begin{cases}
			1 & \text{if } X_t=X^\ast_t \\
			0 & \text{otherwise},
		\end{cases} $$
which we use to express the fact that only individuals in the same vertex may interact and exchange viral load. Here, $X^\ast_t\in\cI$ is the vertex occupied at time $t$ by the other individual involved in the interaction.
\end{enumerate}

To pass from the particle model~\eqref{eq:particle.gen} to an aggregate description, which is more suited to the investigation of the emerging collective trends, we introduce the distribution function of the viral load $v\in\R_+$ at time $t>0$ in vertex $i\in\cI$, that we denote by $f_i=f_i(v,t):\R_+\times\R_+\to\R_+$. Taking advantage of the procedure detailed in~\cite{loy2021KRM_preprint}, see also~\cite{pareschi2013BOOK} for further reference, in the continuous-time limit $\Delta{t}\to 0^+$ we find that $f_i$ satisfies formally the following Boltzmann-type kinetic equation in weak form:
\begin{align}
	\begin{aligned}[b]
		\frac{d}{dt}\int_{\R_+}\varphi(v)f_i(v,t)\,dv &= \chi\int_{\R_+}\varphi(v)\left(\sum_{j\in\cI}P_{ij}f_j(v,t)-f_i(v,t)\right)dv \\
		&\phantom{=} +\mu\int_{\R_+}\varphi(v)Q(f_i)(v,t)\,dv, & i\in\cI,
	\end{aligned}
	\label{eq:boltz.fi}
\end{align}
where $\varphi:\R_+\to\R$ is an arbitrary observable quantity (test function) and $Q(f_i)$ is the so-called \textit{collisional operator} (in the jargon of classical kinetic theory), which here describes the effect of binary interactions among pairs of individuals belonging to the same vertex of the graph. More specifically:
\begin{equation}
	\int_{\R_+}\varphi(v)Q(f_i)(v,t)\,dv=\int_{\R_+}\int_{\R_+}\ave{\varphi(v')-\varphi(v)}f_i(v,t)f_i(v_\ast,t)\,dv\,dv_\ast,
	\label{eq:Q1}
\end{equation}
where $v'$ is given by~\eqref{eq:binary_gen_sym}-\eqref{eq:pq} as a function of the integration variables $v$, $v_\ast$.

\subsection{Aggregate description of mobility and contagion}
\label{sect:aggregate}
We observe that~\eqref{eq:boltz.fi} is a \textit{non-conservative} kinetic equation, indeed the density of individuals in a vertex of the graph, i.e.
$$ \rho_i(t):=\int_{\R_+}f_i(v,t)\,dv\geq 0, $$
is in general not conserved in time because of the jumps across the vertices. Setting $\varphi(v)=1$ in~\eqref{eq:boltz.fi} we obtain that $\rho_i$ satisfies the equation
\begin{equation}
	\frac{d\rho_i}{dt}=\chi\left(\sum_{j\in\cI}P_{ij}\rho_j-\rho_i\right), \qquad i\in\cI,
	\label{eq:rho_i}
\end{equation}
which in vector notation reads
\begin{equation}
	\frac{d\brho}{dt}=\chi(\bP-\bI)\brho
	\label{eq:rho_vect}
\end{equation}
with $\brho:=(\rho_i)_{i\in\cI}$. From~\eqref{eq:rho_vect} we can investigate the stationary mass distribution $\brho^\infty\in\R^n_+$ emerging for large times, namely the vector satisfying the equation
\begin{equation}
	(\bP-\bI)\brho^\infty=\bzero.
	\label{eq:rho.inf}
\end{equation}
In practice, $\brho^\infty$ should be an eigenvector of $\bP$ corresponding to the eigenvalue $1$. Notice that $\bP$ admits indeed this eigenvalue, which coincides also with its spectral radius, because it is a stochastic matrix. Moreover, since $\bP$ is irreducible from Perron-Frobenius theory it follows that the eigenvalue $1$ is simple and that there exists a corresponding eigenvector with strictly positive components. See e.g.,~\cite{minc1988BOOK} for details. Such an eigenvector is our candidate for $\brho^\infty$, but we need to manage its non-uniqueness (scalar multiples are also eigenvectors). For this, we notice that in our case the $\ell^1$-norm of $\brho$, i.e. the quantity $\|\brho\|_1:=\sum_{i\in\cI}\rho_i$, is conserved in time, indeed from~\eqref{eq:rho_i} and~\eqref{eq:P} we deduce
$$ \frac{d}{dt}\|\brho(t)\|_1:=\frac{d}{dt}\sum_{i\in\cI}\rho_i(t)=0. $$
Thus $\|\brho^\infty\|_1$ is fixed by the initial condition, which removes the remaining degree of freedom in the identification of a unique physically admissible $\brho^\infty\in\R^n_+$ solving~\eqref{eq:rho.inf}.

Still from Perron-Frobenius theory we have that $1$ is also the spectral limit of $\bP$, namely the maximum real part of its eigenvalues. Therefore the real parts of the eigenvalues of the matrix $\bP-\bI$ are non-positive and, in particular, the unique eigenvalue with null real part is simple. This implies that $\brho^\infty$ is a stable equilibrium of~\eqref{eq:rho_vect}. Actually, it is also attractive because the eigenvalue with null real part is simply associated with the conservation of the $\ell^1$-norm discussed above.

Summarising, we have proved:
\begin{proposition} \label{prop:rho^inf}
There exists a unique physically admissible solution $\brho^\infty\in\R^n_+$ to~\eqref{eq:rho.inf}, which is a stable and attractive asymptotic density distribution for~\eqref{eq:rho_vect}.
\end{proposition}

Conversely, setting $\varphi(v)=v$ in~\eqref{eq:boltz.fi} we may investigate the evolution of the mean viral load
$$ m_i(t):=\frac{1}{\rho_i(t)}\int_{\R_+}vf_i(v,t)\,dv, $$
which, recalling also~\eqref{eq:binary_gen_sym}-\eqref{eq:pq}, turns out to satisfy the equation
\begin{equation}
	\frac{d}{dt}(\rho_im_i)=\chi\left(\sum_{j\in\cI}P_{ij}\rho_jm_j-\rho_im_i\right)+\mu(\nu_2-\nu_1)\rho_i^2m_i.
	\label{eq:m_i}
\end{equation}
From here, invoking~\eqref{eq:P} we discover
$$ \frac{d}{dt}\sum_{i\in\cI}\rho_im_i=\mu(\nu_2-\nu_1)\sum_{i\in\cI}\rho_i^2m_i, $$
which we may use as a basis to study the large time trend of the $m_i$'s. In particular, we are interested in the stability of the asymptotic state $\bm^\infty=\bzero$, which represents the eradication of the infection in all nodes of the network. Linearising the previous equation around the equilibrium configuration $(\brho,\,\bm)=(\brho^\infty,\,\bzero)$ we obtain
\begin{equation}
	\frac{d}{dt}\sum_{i\in\cI}\rho_i^\infty m_i=\mu(\nu_2-\nu_1)\sum_{i\in\cI}(\rho_i^\infty)^2m_i.
	\label{eq:sum_rhom}
\end{equation}
Next, recalling that $\brho^\infty$ has strictly positive components we set $c:=\min_{i\in\cI}\rho_i^\infty>0$ and we use
$$ \sum_{i\in\cI}(\rho_i^\infty)^2m_i\geq c\sum_{i\in\cI}\rho_i^\infty m_i $$ to conclude that:
\begin{enumerate}[label=\roman*)]
\item if $\nu_1>\nu_2$ then
$$ \frac{d}{dt}\sum_{i\in\cI}\rho_i^\infty m_i\leq c\mu(\nu_2-\nu_1)\sum_{i\in\cI}\rho_i^\infty m_i, $$
which for $t\to+\infty$ implies $\sum_{i\in\cI}\rho_i^\infty m_i\to 0$ and therefore $m_i\to 0$ for all $i\in\cI$. In this case the configuration $\bm^\infty=\bzero$ is locally asymptotically stable, hence the infection may be globally eradicated in the long run;
\item if $\nu_1<\nu_2$ then
$$ \frac{d}{dt}\sum_{i\in\cI}\rho_i^\infty m_i\geq c\mu(\nu_2-\nu_1)\sum_{i\in\cI}\rho_i^\infty m_i, $$
which for $t\to+\infty$ yields $\sum_{i\in\cI}\rho_i^\infty m_i\to+\infty$ and therefore $m_i\to+\infty$ for some $i\in\cI$. In this case the configuration $\bm^\infty=\bzero$ is unstable: the infection is not under control and tends to grow in time;
\item if $\nu_1=\nu_2$ then $\sum_{i\in\cI}\rho_im_i$ is constant in time. This indicates that the configuration $\bm^\infty=\bzero$ is stable but not attractive: the infection is under control but cannot be eradicated and becomes endemic.
\end{enumerate}

\subsubsection{Node-dependent decay and transmission rates and hydrodynamic limit on the network}
If the decay and transmission rates $\nu_1$, $\nu_2$ are not constant but vary from vertex to vertex of the graph, i.e. $\nu_1=\nu_{1,i}$ and $\nu_2=\nu_{2,i}$, then~\eqref{eq:sum_rhom} modifies as
$$ \frac{d}{dt}\sum_{i\in\cI}\rho_i^\infty m_i=\mu\sum_{i\in\cI}(\nu_{2,i}-\nu_{1,i})(\rho_i^\infty)^2m_i. $$
Apart from the relatively trivial cases in which either $\nu_{1,i}>\nu_{2,i}$, $\nu_{1,i}<\nu_{2,i}$ or $\nu_{1,i}=\nu_{2,i}$ for all $i\in\cI$, which actually reproduce the already considered scenarios with constant rates, this equation is not particularly informative about the large time trend of the infection. We may get more detailed information by looking instead at the \textit{hydrodynamic regime} of the kinetic equation~\eqref{eq:boltz.fi}, namely the one in which local interactions within the vertices of the graph are much more frequent than jumps from node to node. This amounts to scaling
$$ \chi=1, \qquad \mu=\frac{1}{\epsilon}, $$
where $0<\epsilon\ll 1$ is a small parameter playing conceptually the role of the Knudsen number of classical kinetic theory. In the hydrodynamic limit $\epsilon\to 0^+$ a splitting of~\eqref{eq:boltz.fi} is possible in:
\begin{enumerate}[label=\roman*)]
\item intra-vertex interactions without jumps on the quick time scale $\tau:=t/\epsilon$, ruled by
\begin{equation}
	\frac{d}{d\tau}\int_{\R_+}\varphi(v)f_i(v,\tau)\,dv=\int_{\R_+}\int_{\R_+}\ave{\varphi(v')-\varphi(v)}f_i(v,\tau)f_i(v_\ast,\tau)\,dv\,dv_\ast,
	\label{eq:split.int}
\end{equation}
which lead $f_i$ to converge for large $\tau$ to a local equilibrium distribution $M_i$ (the \textit{local Maxwellian}) parametrised by the macroscopic quantities conserved by the interactions (the \textit{collisional invariants}) in the $i$th vertex;
\item jumps across the vertices without interactions on the slow time scale $t$, ruled by
\begin{equation}
	\frac{d}{dt}\int_{\R_+}\varphi(v)M_i(v,t)\,dv=\int_{\R_+}\varphi(v)\left(\sum_{j\in\cI}P_{ij}M_j(v,t)-M_i(v,t)\right)dv,
	\label{eq:split.jump}
\end{equation}
which determine the macroscopic evolution of the quantities conserved by the local intra-vertex interactions.
\end{enumerate}
Taking $\varphi(v)=1$ in~\eqref{eq:split.int} we discover that in all vertices the density $\rho_i$ is constant on the time scale $\tau$. Taking instead $\varphi(v)=v$ we get
\begin{equation}
	\frac{dm_i}{d\tau}=(\nu_{2,i}-\nu_{1,i})\rho_im_i.
	\label{eq:mi}
\end{equation}

Let us assume first that $\nu_{1,i}>\nu_{2,i}$ for all $i\in\cI\setminus\{i^\ast\}$ while $\nu_{1,i^\ast}=\nu_{2,i^\ast}$. Then~\eqref{eq:mi} implies $m_i\to 0$ exponentially fast for $\tau\to+\infty$ if $i\neq i^\ast$ while $m_{i^\ast}$ is constant on the $\tau$-scale. Thus for $i\neq i^\ast$ the local Maxwellian is $M_{\rho_i}(v)=\rho_i\delta(v)$, $\delta$ denoting the Dirac delta centred in $v=0$, whereas for $i=i^\ast$ the local Maxwellian is a distribution $M_{\rho_{i^\ast},m_{i^\ast}}(v)$ parametrised by both $\rho_{i^\ast}$ and $m_{i^\ast}$. Plugging these local Maxwellians into~\eqref{eq:split.jump} and choosing $\varphi(v)=1$ yields the same time evolution of the densities $\rho_i$, $i\in\cI$, as~\eqref{eq:rho_i}. Then, in particular, $\rho_i\to\rho^\infty_i>0$ for all $i\in\cI$ when $t\to +\infty$ as asserted by Proposition~\ref{prop:rho^inf}. Choosing instead $\varphi(v)=v$ and $i=i^\ast$ we may investigate the evolution of the mean viral load in the $i^\ast$th vertex, the only one which remains constant during the intra-vertex interactions:
$$ \frac{d}{dt}(\rho_{i^\ast}m_{i^\ast})=(P_{i^\ast i^\ast}-1)\rho_{i^\ast}m_{i^\ast}. $$
Since the graph is strongly connected we have in particular $P_{i^\ast i^\ast}<1$, whence $\rho_{i^\ast}m_{i^\ast}\to 0$ for $t\to+\infty$. Considering that $\rho_{i^\ast}$ converges to a non-zero asymptotic value, we finally infer $m_{i^\ast}\to 0$. Thus, in the long run, thanks to the mobility of the individuals across the nodes of the network the infection is eradicated also in the node where local contagion dynamics balance with the physiological recovery. Notice however that such an eradication is slower than in the other nodes, indeed it happens on the $t$-time scale rather than on the quicker $\tau$-scale.

Next, let us consider the opposite situation, namely $\nu_{1,i}=\nu_{2,i}$ for all $i\in\cI\setminus\{i^\ast\}$ and $\nu_{1,i^\ast}>\nu_{2,i^\ast}$. In this case, in all vertices $i\neq i^\ast$ the local contagion dynamics yield local Maxwellians $M_{\rho_i,m_i}(v)$ parametrised by both $\rho_i$ and $m_i$ while in the $i^\ast$th vertex it results $m_{i^\ast}\to 0$ as $\tau\to+\infty$, thus the local Maxwellian is $M_{\rho_{i^\ast}}(v)=\rho_{i^\ast}\delta(v)$. Consequently from~\eqref{eq:split.jump} with $\varphi(v)=v$ we deduce
$$ \frac{d}{dt}(\rho_im_i)=\sum_{j\in\cI\setminus\{i^\ast\}}P_{ij}\rho_jm_j-\rho_im_i, \qquad i\in\cI\setminus\{i^\ast\}, $$
whence, summing both sides over $i\in\cI\setminus\{i^\ast\}$,
\begin{align}
	\begin{aligned}[b]
		\frac{d}{dt}\sum_{i\in\cI\setminus\{i^\ast\}}\rho_im_i &= \sum_{j\in\cI\setminus\{i^\ast\}}\left(\sum_{i\in\cI\setminus\{i^\ast\}}P_{ij}\right)\rho_jm_j
			-\sum_{i\in\cI\setminus\{i^\ast\}}\rho_im_i \\
		&= -\sum_{j\in\cI\setminus\{i^\ast\}}P_{i^\ast j}\rho_jm_j,
	\end{aligned}
	\label{eq:rhoi.mi.iast-sum}
\end{align}
where we have used the fact that $\sum_{i\in\cI\setminus\{i^\ast\}}P_{ij}=1-P_{i^\ast j}$. Let us introduce the set of indices of the vertices different from vertex $i^\ast$ and directly linked to the latter:
$$ \cI^\ast:=\{j\in\cI\setminus\{i^\ast\}\,:\,P_{i^\ast j}>0\}. $$
Notice that $\cI^\ast$ is non-empty due to the strong connectivity of the graph (if $P_{i^\ast j}=0$ for all $j\neq i^\ast$ then vertex $i^\ast$ could not be reached from any other vertex). Let $a:=\min_{j\in\cI^\ast}P_{i^\ast j}>0$, then from~\eqref{eq:rhoi.mi.iast-sum} we deduce
$$ \frac{d}{dt}\sum_{i\in\cI\setminus\{i^\ast\}}\rho_im_i\leq -a\sum_{j\in\cI^\ast}\rho_jm_j, $$
whence, integrating both sides in time and considering that $\sum_{i\in\cI\setminus\{i^\ast\}}\rho_im_i\geq \sum_{i\in\cI^\ast}\rho_im_i$ because $\cI^\ast\subseteq\cI\setminus\{i^\ast\}$,
$$ \sum_{i\in\cI^\ast}\rho_i(t)m_i(t)\leq \sum_{i\in\cI\setminus\{i^\ast\}}\rho_{i,0}m_{i,0}-a\int_0^t\sum_{j\in\cI^\ast}\rho_j(s)m_j(s)\,ds. $$
By Gr\"{o}nwall's inequality we deduce then
$$ \sum_{i\in\cI^\ast}\rho_i(t)m_i(t)\leq e^{-at}\sum_{i\in\cI\setminus\{i^\ast\}}\rho_{i,0}m_{i,0}, $$
which says that $\sum_{i\in\cI^\ast}\rho_im_i\to 0$, hence that $m_i\to 0$ for all $i\in\cI^\ast$ as $t\to+\infty$. We conclude that, owing to the mobility of the individuals, the infection is certainly eradicated at least in the vertices directly linked to the one where contagion dynamics are weaker than the physiological recovery of the individuals. In particular, if $\cI^\ast=\cI\setminus\{i^\ast\}$, i.e. if all vertices of the graph are directly linked to vertex $i^\ast$, then the infection is eradicated in the whole network. It is worth stressing that instead in classical homogeneous kinetic models featuring binary interactions of the form~\eqref{eq:binary_gen_sym} with coefficients given by~\eqref{eq:pq} there is no way to obtain a convergence to zero of the mean if $\nu_1=\nu_2$.

Finally, if $\nu_{1,i^\ast}<\nu_{2,i^\ast}$ for some $i^\ast\in\cI$ then from~\eqref{eq:mi} we easily compute that the intra-vertex interactions produce locally a blow-up of the mean viral load: $m_{i^\ast}\to+\infty$ as $\tau\to+\infty$. The mean viral load will consequently blow in every other vertex $i\in\cI$ where $\nu_{1,i}\leq\nu_{2,i}$ while the infection will be eradicated only in the vertices where $\nu_{1,i}>\nu_{2,i}$.

\subsection{The problem of commuters}
\label{sect:commuters}
A particularly interesting variation on the model introduced in the previous sections is the case of commuters, namely individuals who do not travel generically between any two vertices of the graph but mainly between two specific vertices representing e.g., the places where they live and work. For the sake of simplicity, we will assume that they travel only between those two vertices.

A possible particle description of the problem is as follows. Let $(X_{o,t},\,X_{d,t})\in\cI\times\cI$ be the origin-destination pair of vertices of a commuter who at time $t$ is in $X_{o,t}$ and is heading for $X_{d,t}$. Let moreover $V_t\in\R_+$ be their viral load. We model the evolution of $(X_{o,t},\,X_{d,t})$ and $V_t$ during a time step $\Delta{t}>0$ as
\begin{align}
	\begin{aligned}[c]
		(X_{o,t+\Delta{t}},\,X_{d,t+\Delta{t}}) &= (1-\Theta)(X_{o,t},\,X_{d,t})+\Theta(X'_{o,t},\,X'_{d,t}) \\
		V_{t+\Delta{t}} &= \left(1-\Xi\delta_{X_{o,t},X^\ast_{o,t}}\right)V_t+\Xi\delta_{X_{o,t},X^\ast_{o,t}}V'_t,
	\end{aligned}
	\label{eq:particle.commuters}
\end{align}
where $\Theta$, $\Xi$ are like in~\eqref{eq:Theta.Xi}, $V'_t=pV_t+qV^\ast_t$ from~\eqref{eq:binary_gen_sym} and $V^\ast_t$, $X^\ast_{o,t}$ are the viral load and the origin vertex of the other individual participating in the interaction. The vector $(X'_{o,t},\,X'_{d,t})\in\cI\times\cI$ expresses the new origin-destination pair of vertices of the commuter after a possible journey from $X_{o,t}$ to $X_{d,t}$. Notice that, by definition of commuter, we may only have either $X'_{o,t}=X_{d,t}$ and $X'_{d,t}=X_{o,t}$ if the journey takes place or $X'_{o,t}=X_{o,t}$ and $X'_{d,t}=X_{d,t}$ if it does not. Therefore, if we denote by $p_{X_{d,t},X_{o,t}}\in [0,\,1]$ the probability for such a commuter to actually travel from $X_{o,t}$ to $X_{d,t}$ then the law of the random vector $(X'_{o,t},\,X'_{d,t})$ is
\begin{align*}
	\Prob((X'_{o,t},\,X'_{d,t})=(X_{d,t},\,X_{o,t})) &= p_{X_{d,t},X_{o,t}}, \\
	\Prob((X'_{o,t},\,X'_{d,t})=(X_{o,t},\,X_{d,t})) &= 1-p_{X_{d,t},X_{o,t}}.
\end{align*}

We are now in a position to deduce from~\eqref{eq:particle.commuters} an aggregate description in terms of the distribution function of the viral load $v\in\R_+$ of commuters between vertices $i,\,j\in\cI$ (in the following we will refer to them as $ij$-commuters for brevity) who at time $t>0$ are in vertex $i$ and are heading for vertex $j$. Let us denote by $f_{ji}=f_{ji}(v,t):\R_+\times\R_+\to\R_+$ such a distribution function. Using again the technique illustrated in~\cite{loy2021KRM_preprint} with minimal adaptations, in the continuous-time limit $\Delta{t}\to 0^+$ we formally obtain from~\eqref{eq:particle.commuters} that $f_{ji}$ satisfies the following Boltzmann-type kinetic equation in weak form:
\begin{align}
	\begin{aligned}[b]
		\frac{d}{dt}\int_{\R_+}\varphi(v)f_{ji}(v,t)\,dv &= \chi\int_{\R_+}\varphi(v)\left(p_{ij}f_{ij}(v,t)-p_{ji}f_{ji}(v,t)\right)dv \\
		&\phantom{=} +\mu\int_{\R_+}\int_{\R_+}\ave{\varphi(v')-\varphi(v)}f_{ji}(v,t)f_i(v_\ast,t)\,dv\,dv_\ast, & i,\,j\in\cI,
	\end{aligned}
	\label{eq:boltz.commuters}
\end{align}
where $v'$ is given by~\eqref{eq:binary_gen_sym}-\eqref{eq:pq} and
$$ f_i(v,t):=\sum_{j\in\cI}f_{ji}(v,t) $$
is the cumulative distribution function of the viral load of the individuals who at time $t$ are in vertex $i$ (independently of the other vertex to which they commute).

Equation~\eqref{eq:boltz.commuters} features two important differences with respect to~\eqref{eq:boltz.fi}, that we now elucidate.
\begin{enumerate}[label=\roman*)]
\item The commuting term
$$ \int_{\R_+}\varphi(v)\left(p_{ij}f_{ij}(v,t)-p_{ji}f_{ji}(v,t)\right)dv $$
allows only for exchanges between $ij$-commuters in $j$ heading for $i$ ($p_{ij}f_{ij}$, gain) and $ij$-commuters in $i$ heading for $j$ ($p_{ji}f_{ji}$, loss). Notice indeed that, by definition of $ij$-commuters, individuals arriving in vertex $i$ from vertices different from $j$ do not contribute to the time variation of $f_{ji}$.
\item The contagion term
$$ \int_{\R_+}\int_{\R_+}\ave{\varphi(v')-\varphi(v)}f_{ji}(v,t)f_i(v_\ast,t)\,dv\,dv_\ast $$
takes into account binary interactions between $ij$-commuters in $i$ and all other individuals in vertex $i$, independently of the vertex to which they commute, because they all contribute equally to the spread of the contagion in vertex $i$.
\end{enumerate}

\begin{remark}
We stress that, in general, $p_{ij}\neq P_{ij}$, cf.~\eqref{eq:Pij}. Indeed, in the model introduced in Section~\ref{sect:Boltzmann.graph}, $P_{ij}$ is the probability that \textit{any} individual in vertex $j$ jumps to vertex $i$ while here $p_{ij}$ is the probability that \textit{an $ij$-commuter} currently in vertex $j$ jumps to vertex $i$ (all individuals in vertex $j$ might not be $ij$-commuters).
\label{rem:commuters}
\end{remark}

If we define by
$$ \rho_{ji}(t):=\int_{\R_+}f_{ji}(v,t)\,dv $$
the density of $ij$-commuters in vertex $i$ at time $t$ then from~\eqref{eq:boltz.commuters} with $\varphi(v)=1$ we obtain the macroscopic equation of the mobility on the graph:
\begin{equation}
	\frac{d\rho_{ji}}{dt}=\chi\left(p_{ij}\rho_{ij}-p_{ji}\rho_{ji}\right), \qquad i,\,j\in\cI.
	\label{eq:rhoji}
\end{equation}
Notice that the time evolution of $\rho_{ji}$ depends only on $\rho_{ji}$ itself and on $\rho_{ij}$, which is the density of $ij$-commuters travelling in the opposite direction. This is consistent with the definition of commuters. Moreover, from the analogous equation for $\rho_{ij}$ we easily see that the total density of $ij$-commuters, namely $\rho_{ji}+\rho_{ij}$, is constant in time. Let us denote by\footnote{We use the set-style subscript $\{i,j\}$ to mean that, unlike $f_{ji}$ and $\rho_{ji}$, in the notation $\bar{\rho}_{\{i,j\}}$ there is no origin-destination ordering of the indices $i$, $j$ because $\bar{\rho}_{\{i,j\}}$ is the total mass of $ij$-commuters independently of the direction of their journey.} $\bar{\rho}_{\{i,j\}}:=\rho_{ji}+\rho_{ij}$ the total mass of $ij$-commuters. Then $\rho_{ij}=\bar{\rho}_{\{i,j\}}-\rho_{ji}$, whence
$$ \frac{d\rho_{ji}}{dt}=\chi\left(p_{ij}\bar{\rho}_{\{i,j\}}-(p_{ji}+p_{ij})\rho_{ji}\right), $$
which implies that $\rho_{ji}$, $\rho_{ij}$ reach exponentially fast the asymptotic values
$$ \rho^\infty_{ji}:=\frac{p_{ij}}{p_{ji}+p_{ij}}\bar{\rho}_{\{i,j\}}, \qquad \rho^\infty_{ij}:=\frac{p_{ji}}{p_{ji}+p_{ij}}\bar{\rho}_{\{i,j\}}, \qquad i,\,j\in\cI. $$
From this, summing over all destination vertices, we may also compute the asymptotic value of
$$ \rho_i(t):=\sum_{j\in\cI}\rho_{ji}(t)=\int_{\R_+}f_i(v,t)\,dv, $$
i.e. the total density of individuals in vertex $i$ at time $t$. We have:
\begin{equation}
	\rho^\infty_i=\sum_{j\in\cI}\frac{p_{ij}}{p_{ji}+p_{ij}}\bar{\rho}_{\{i,j\}}, \qquad i\in\cI.
	\label{eq:rho^inf_i}
\end{equation}

\begin{remark}
From~\eqref{eq:rhoji}, the evolution equation for the total density $\rho_i$ turns out to be
$$ \frac{d\rho_i}{dt}=\chi\sum_{j\in\cI}\left(p_{ij}\rho_{ij}-p_{ji}\rho_{ji}\right), $$
which is here the counterpart of~\eqref{eq:rho_i}. If we assume that all commuters always travel from their origin to their destination, i.e. that no commuter might remain in the origin vertex without travelling, then we have $p_{ji}=1$ for all $i,\,j\in\cI$ and this equation becomes
$$ \frac{d\rho_i}{dt}=\chi\left(\sum_{j\in\cI}\rho_{ij}-\rho_i\right), \qquad i\in\cI, $$
which reminds more closely of~\eqref{eq:rho_i} but with $P_{ij}\rho_j$ replaced by $\rho_{ij}$. Notice that, unlike~\eqref{eq:rho_i}, this is not a self-consistent equation in terms of the $\rho_i$'s alone because commuting requires to keep track of origin and destination vertices.
\end{remark}

Likewise, if we define by
$$ m_{ji}(t):=\frac{1}{\rho_{ji}(t)}\int_{\R_+}vf_{ji}(v,t)\,dv $$
the mean viral load of $ij$-commuters in vertex $i$ at time $t$ then from~\eqref{eq:boltz.commuters} with $\varphi(v)=v$ and taking also~\eqref{eq:binary_gen_sym}-\eqref{eq:pq} and~\eqref{eq:rhoji} into account we obtain
\begin{equation}
	\frac{dm_{ji}}{dt}=\chi p_{ij}\frac{\rho_{ij}}{\rho_{ji}}(m_{ij}-m_{ji})+\mu\rho_i\left(\nu_2m_i-\nu_1m_{ji}\right),
		\qquad i,\,j\in\cI,
	\label{eq:mji}
\end{equation}
where
$$ m_i(t):=\frac{1}{\rho_i(t)}\int_{\R_+}vf_i(v,t)\,dv=\frac{1}{\rho_i(t)}\sum_{j\in\cI}\rho_{ji}(t)m_{ji}(t) $$
is the total mean viral load in vertex $i$. This term couples macroscopically the contagion dynamics of $ij$-commuters to those of all commuters in $i$ travelling to other vertices different from $j$. This reflects the fact that contagion dynamics are not confined to the transport of viral load along the commuting routes (cf. the first term on the right-hand side of~\eqref{eq:mji}). Contagion may spread along all routes because of mixed social contacts in each vertex.

\section{Social interactions with quarantine in the nodes}
\label{sect:quarantine}
We consider now the case in which individuals may be quarantined in each node of the network if they are diagnosed as infected. Quarantined individuals do not have social contacts with other individuals, therefore they only diminish their viral load and are eventually readmitted in the society when they are diagnosed as recovered by a subsequent viral test.

To approach this problem we take advantage of the framework introduced in~\cite{loy2021KRM_preprint}, where kinetic equations with \textit{label switching} are discussed. In essence, in each vertex $i\in\cI$ of the graph we label the individuals by $h=1$ if they are not quarantined and by $h=2$ if instead they are. Next, we introduce the \textit{label switch probability}
$$ T^{hk}_i(v):=\Prob(k\to h\vert i,\,v)\in [0,\,1], \qquad h,\,k\in\{1,\,2\},\ i\in\cI,\ v\in\R_+, $$
namely the probability that an individual in vertex $i$ with viral load $v$ changes their label from $k$ to $h$. In particular, the label switching $1\to 2$ means that the individual is quarantined; conversely, the label switching $2\to 1$ means that the individual is released from quarantine. When $h=k$ the individual does not change their quarantine state. These probabilities fulfil
\begin{equation}
	T^{1k}_i(v)+T^{2k}_i(v)=1 \qquad \forall\,k\in\{1,\,2\},\ i\in\cI,\ v\in\R_+.
	\label{eq:sum.Tihk}
\end{equation}

From the particle point of view, we now characterise the state of a representative individual at time $t$ by the vertex $X_t\in\cI$ to which they belong, the label $H_t\in\{1,\,2\}$ identifying their quarantine state and their viral load $V_t\in\R_+$. Next, we model the state update in a time step $\Delta{t}>0$ as
\begin{align}
	\begin{aligned}[c]
		X_{t+\Delta{t}} &= (1-\Theta)X_t+\Theta J_t \\
		H_{t+\Delta{t}} &= (1-\Lambda)H_t+\Lambda I_t \\
		V_{t+\Delta{t}} &= \left(1-\Xi\right)V_t+\Xi V'_t,
	\end{aligned}
	\label{eq:particle.quarantine}
\end{align}
where $\Lambda\in\{0,\,1\}$ is a Bernoulli random variable such that
$$ \Prob(\Lambda=1)=\lambda\Delta{t} $$
which discriminates whether a label switching takes place ($\Lambda=1$) or not ($\Lambda=0$) in the time step $\Delta{t}$ and $\lambda>0$ is the rate of viral testing. Moreover, $I_t\in\{1,\,2\}$ is the new label assigned to the individual after a viral test, with in particular
$$ \Prob(I_t=h\vert H_t=k,\,X_t=i,\,V_t=v)=T^{hk}_i(v). $$
The random variables $\Theta$, $\Xi$ are like in Section~\ref{sect:Boltzmann.graph} but in this case we fix $\Delta{t}\leq\min\{\frac{1}{\chi},\,\frac{1}{\lambda},\,\frac{1}{\mu}\}$ for consistency. A remarkable difference with respect to the model in Section~\ref{sect:Boltzmann.graph} is instead the structure of the random variables $J_t$, $V'_t$ for now only a non-quarantined individual ($h=1$) can jump to another vertex and can experience social contacts producing viral load changes according to formula~\eqref{eq:binary_gen_sym}. To take these facts into account we modify the law of $J_t$ as follows:
$$ \Prob(J_t=i\vert X_t=j,\,H_t=h)=P^{h}_{ij}\in [0,\,1], $$
where the $P^{1}_{ij}$'s coincide with the $P_{ij}$'s of Section~\ref{sect:Boltzmann.graph} while
$$ P^2_{ij}=\delta_{ij}, \qquad \forall\ i,\,j\in\cI $$
because a quarantined individual cannot change vertex. Parallelly, for quarantined ($h=2$) individuals we convert the binary interaction rule~\eqref{eq:binary_gen_sym}-\eqref{eq:pq} into a rule expressing a progressive reduction of viral load due to the lack of social contacts:
\begin{equation}
	v'=(1-\gamma+\xi)v,
	\label{eq:binary.quarantined}
\end{equation}
which is of the form~\eqref{eq:binary_gen_sym} with $p=1-\gamma+\xi$ and $q=0$. Here, $\gamma\in [0,\,1]$ is the decay rate of the viral load of a quarantined individual, which in general is expected to satisfy $\gamma\geq\nu_1$ because quarantined individuals may be exposed to specific medical treatments in addition to the physiological recovery. Moreover, $\xi\in\R$ is a centred random coefficient modelling stochastic fluctuations of the decay rate. With the restriction $\xi\geq\gamma-1$ we guarantee $v'\geq 0$ for all $v\geq 0$. On the whole, we set
\begin{equation*}
	V'_t=
		\begin{cases}
			(1-\nu_1+\eta)V_t+\nu_2V^\ast_t & \text{if } H_t=H^\ast_t=1 \text{ and } X_t=X^\ast_t \\
			V_t & \text{if } (H_t=H^\ast_t=1 \text{ and } X_t\neq X^\ast_t) \text{ or } (H_t=1 \text{ and } H^\ast_t=2) \\
			(1-\gamma+\xi)V_t & \text{if } H_t=2,
		\end{cases}
\end{equation*}
%\begin{align*}
%	V'_t &= \delta_{H_t,1}\Bigl\{\delta_{H^\ast_t,1}\delta_{X_t,X^\ast_t}[(1-\nu_1+\eta)V_t+\nu_2V^\ast_t]+[\delta_{H^\ast_t,1}(1-\delta_{X_t,X^\ast_t})+\delta_{H^\ast_t,2}]V_t\Bigr\} \\
%	&\phantom{=} +\delta_{H_t,2}(1-\gamma+\xi)V_t,
%\end{align*}
where $H^\ast_t\in\{1,\,2\}$ denotes the label identifying the quarantine state of the other individual participating in the interaction.

Let now $f^h_i=f^h_i(v,t):\R_+\times\R_+\to\R_+$ be the distribution function of the viral load $v\in\R_+$ at time $t>0$ in the vertex $i\in\cI$ for an individual with label $h\in\{1,\,2\}$. Combining~\eqref{eq:boltz.fi} and the results in~\cite{loy2021KRM_preprint} we deduce from~\eqref{eq:particle.quarantine} that in the continuous-time limit $\Delta{t}\to 0^+$ the $f^h_i$'s satisfy formally the following kinetic equation in weak form:
\begin{align}
	\begin{aligned}[b]
		\frac{d}{dt}\int_{\R_+}\varphi(v)f^h_i(v,t)\,dv &= \chi\int_{\R_+}\varphi(v)\left(\sum_{j\in\cI}P^{h}_{ij}f^h_j(v,t)-f^h_i(v,t)\right)dv \\
		&\phantom{=} +\lambda\int_{\R_+}\varphi(v)\left(\sum_{k=1}^{2}T^{hk}_i(v)f^k_i(v,t)-f^h_i(v,t)\right)dv \\
		&\phantom{=} +\mu\int_{\R_+}\varphi(v)Q^h(f^h_i)(v,t)\,dv, & i\in\cI,\,h\in\{1,\,2\},
	\end{aligned}
	\label{eq:bolz.fih}
\end{align}
where the collisional operator $Q^h$ is labelled by $h$ to reflect the different rules of change of viral load followed by non-quarantined and quarantined individuals. Specifically, the collisional operator $Q^1$ coincides with the bilinear one given in~\eqref{eq:Q1} whereas $Q^2$ takes the form of a linear collision operator, i.e. it depends on the term $f^h_i(v,t)$ alone rather than on the product $f^h_i(v,t)f^h_i(v_\ast,t)$. Explicitly, the equations for the two compartments $h=1,\,2$ read then:
\begin{itemize}
\item non-quarantined individuals ($h=1$):
\begin{align}
	\begin{aligned}[b]
		\frac{d}{dt}\int_{\R_+}\varphi(v)f^1_i(v,t)\,dv &= \chi\int_{\R_+}\varphi(v)\left(\sum_{j\in\cI}P^1_{ij}f^1_j(v,t)-f^1_i(v,t)\right)dv \\
		&\phantom{=} +\lambda\int_{\R_+}\varphi(v)\left(\sum_{k=1}^{2}T^{1k}_i(v)f^k_i(v,t)-f^1_i(v,t)\right)dv \\
		&\phantom{=} +\mu\int_{\R_+}\int_{\R_+}\ave{\varphi(v')-\varphi(v)}f^1_i(v,t)f^1_i(v_\ast,t)\,dv\,dv_\ast, & i\in\cI
	\end{aligned}
	\label{eq:bolz.non-quarantined}
\end{align}
with $v'$ given by~\eqref{eq:binary_gen_sym}-\eqref{eq:pq};
\item quarantined individuals ($h=2$):
\begin{align}
	\begin{aligned}[b]
		\frac{d}{dt}\int_{\R_+}\varphi(v)f^2_i(v,t)\,dv &= \lambda\int_{\R_+}\varphi(v)\left(\sum_{k=1}^{2}T^{2k}_i(v)f^k_i(v,t)-f^2_i(v,t)\right)dv \\
		&\phantom{=} +\mu\int_{\R_+}\ave{\varphi(v')-\varphi(v)}f^2_i(v,t)\,dv, & i\in\cI
	\end{aligned}
	\label{eq:bolz.quarantined}
\end{align}
with $v'$ given by~\eqref{eq:binary.quarantined}.
\end{itemize}

\subsection{Hydrodynamic model with quarantine on the network}
We observe that~\eqref{eq:bolz.fih} is a fully non-conservative kinetic equation, indeed it does not even conserve the mass of the individuals belonging to each compartment. Let
$$ \rho^h_i(t):=\int_{\R_+}f^h_i(v,t)\,dv $$
be the density of individuals in vertex $i$ and compartment $h$ at time $t$. Setting $\varphi(v)=1$ in~\eqref{eq:bolz.fih} and observing that $\int_{\R_+}Q^h(f^h_i)(v,t)\,dv=0$ for $h=1,\,2$ because both microscopic rules~\eqref{eq:binary_gen_sym}-\eqref{eq:pq} and~\eqref{eq:binary.quarantined} conserve the number of individuals, we obtain that $\rho^h_i$ satisfies the equation
\begin{equation}
	\frac{d\rho^h_i}{dt}=\chi\left(\sum_{j\in\cI}P^h_{ij}\rho^h_j-\rho^h_i\right)+\lambda\left(\sum_{k=1}^{2}\int_{\R_+}T^{hk}_i(v)f^k_i(v,t)\,dv-\rho^h_i\right),
	\label{eq:rhoih}
\end{equation}
which depends explicitly on the intra-vertex label switching and the jumps from vertex to vertex.

\begin{remark} \label{rem:rhoi}
Summing~\eqref{eq:rhoih} over $h$ and defining the global density of individuals in vertex $i$ at time $t$ as
$$ \rho_i(t):=\sum_{h=1}^{2}\rho^h_i(t) $$
we find
$$ \frac{d\rho_i}{dt}=\chi\left(\sum_{j\in\cI}\sum_{h=1}^{2}P^h_{ij}\rho^h_j-\rho_i\right), $$
where we have recalled~\eqref{eq:sum.Tihk}. If the transition probabilities from vertex to vertex do not depend on $h$ then this equation coincides with~\eqref{eq:rho_i}. Otherwise it describes different aggregate dynamics, that cannot be expressed in terms of the $\rho_i$'s alone. This is the case of the model with quarantine, as we have in general $P^1_{ij}\neq P^2_{ij}$ due to different mobility features of the individuals in the two compartments $h=1,\,2$.
\end{remark}

Remark~\ref{rem:rhoi} implies that~\eqref{eq:rhoih} is the reference equation for the evolution of the density of individuals on the network in the model with quarantine. Notice that it requires, in general, the knowledge of the kinetic distribution functions $f^k_i$ to compute the label switching at the macroscopic scale. Nevertheless, in the particular case that the label switch probabilities $T^{hk}_i$ are constant with respect to $v$, namely if we assume that the probability to diagnose an individual as either infected or recovered may be considered approximately independent of their viral load,~\eqref{eq:rhoih} becomes instead a self-consistent equation for the densities $\rho^h_i$:
$$ \frac{d\rho^h_i}{dt}=\chi\left(\sum_{j\in\cI}P^h_{ij}\rho^h_j-\rho_i\right)+\lambda\left(\sum_{k=1}^{2}T^{hk}_i\rho^k_i-\rho^h_i\right), $$
which for $h=1,\,2$ and recalling~\eqref{eq:sum.Tihk} produces the system
\begin{equation}
	\begin{cases}
		\dfrac{d\rho^1_i}{dt}=\chi\left(\displaystyle{\sum_{j\in\cI}}P^1_{ij}\rho^1_j-\rho^1_i\right)+\lambda\left(T^{12}_i\rho^2_i-T^{21}_i\rho^1_i\right) \\[5mm]
		\dfrac{d\rho^2_i}{dt}=\lambda\left(T^{21}_i\rho^1_i-T^{12}_i\rho^2_i\right),
	\end{cases}
	\quad
	i\in\cI.
	\label{eq:rhoih.system}
\end{equation}

System~\eqref{eq:rhoih.system} provides self-consistent macroscopic information on the mass distribution of the individuals on the network but not specifically on the aggregate trends of the infection itself. To get a more complete picture of the epidemic spread, we may use~\eqref{eq:bolz.non-quarantined}-\eqref{eq:bolz.quarantined} to deduce a coupled system of hydrodynamic equations for the mean viral loads
$$ m^h_i(t):=\frac{1}{\rho^h_i(t)}\int_{\R_+}vf^h_i(v,t)\,dv, \qquad i\in\cI,\,h\in\{1,\,2\}. $$
Letting $\varphi(v)=v$ in~\eqref{eq:bolz.non-quarantined}-\eqref{eq:bolz.quarantined} and assuming again that the $T^{hk}_i$'s are constant with respect to $v$ we obtain, after some manipulations taking also advantage of~\eqref{eq:rhoih.system},
\begin{equation}
	\begin{cases}
		\dfrac{dm^1_i}{dt}=\chi\displaystyle{\sum_{j\in\cI}}P^1_{ij}\dfrac{\rho^1_j}{\rho^1_i}\left(m^1_j-m^1_i\right)
			+\lambda T^{12}_i\dfrac{\rho^2_i}{\rho^1_i}\left(m^2_i-m^1_i\right)+\mu(\nu_2-\nu_1)\rho^1_im^1_i \\[5mm]
		\dfrac{dm^2_i}{dt}=\lambda T^{21}_i\dfrac{\rho^1_i}{\rho^2_i}\left(m^1_i-m^2_i\right)-\mu\gamma m^2_i,
	\end{cases}
	\quad
	i\in\cI.
	\label{eq:mih.system}
\end{equation}

It is now more difficult to extract from~\eqref{eq:rhoih.system}-\eqref{eq:mih.system} analytical information on the large time trends of the model and its equilibria due to the substantial lack of basic conservation properties. In Section~\ref{sect:numerics} we will explore numerically some representative case studies. Furthermore, we mention that in~\cite{loy2021KRM_preprint} a qualitative analysis is proposed concerning the aggregate time asymptotic trends of the quarantine model without network (i.e. formally $\abs{\cI}=1$) for both constant and non-constant label switch probabilities.

\section{Numerical tests}
\label{sect:numerics}
\begin{table}[!t]
\centering
\caption{Parameters of the interactions used in the numerical tests of Section~\ref{sect:numerics}. The symbol $\mathcal{U}([-a,\,a])$ indicates a random variable uniformly distributed in the interval $[-a,\,a]\subset\R$, $a>0$}
\label{tab:param}
\begin{tabular}{c|cccccc}
\hline
\multirow{2}{*}{Parameter} & Test 1 & Test 2 & Test 3 \\
& (Figure~\ref{fig:net_int}) & (Figure~\ref{fig:commuters}) & (Figures~\ref{fig:quarantine-cp0},~\ref{fig:quarantine-cp1}) \\
\hline
\hline
$\chi$ & $1$ & $1$ & $1$ \\
$\mu$ & $1$ & $1$ & $1$ \\
$\lambda$ & -- & -- & $1$ \\
$\nu_1$ & $0$ & $0$ & $0$ \\
$\nu_2$ & $0.2$ & $0.2$ & $0.2$ \\
$\eta$ & $\mathcal{U}([-1,\,1])$ & $\mathcal{U}([-1,\,1])$ & $\mathcal{U}([-1,\,1])$ \\
$\gamma$ & -- & -- & $0.3$\\
$\xi$ & -- & -- & $\mathcal{U}([-0.7,\,0.7])$ \\
\hline
\end{tabular}
\end{table}

\begin{figure}[!t]
\centering
\includegraphics[width=.7\textwidth]{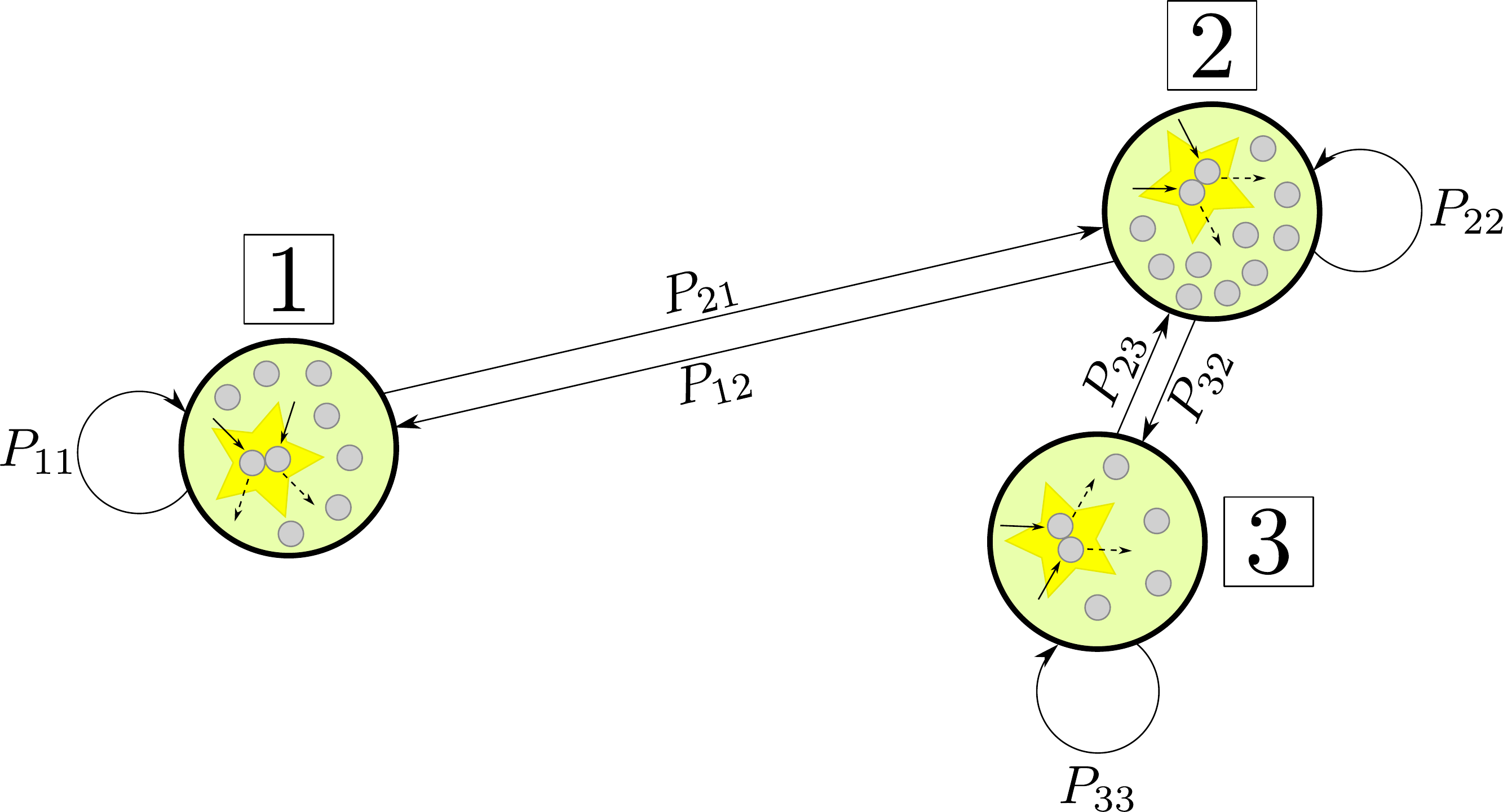}
\caption{The three-node network considered in the numerical tests of Section~\ref{sect:numerics}}
\label{fig:network3nodes}
\end{figure}

We present now some illustrative numerical tests, which exemplify the type of results and considerations that can be drawn from the models introduced in the previous sections. In each test, we compute the aggregate densities of individuals and the mean viral loads by solving both the particle model via suitable adaptations of classical Monte Carlo schemes for kinetic equations, cf. e.g.,~\cite{pareschi2013BOOK}, and the macroscopic equations via standard numerical methods for ODEs such as the fourth-order Runge-Kutta method. Compliance of the two solutions validates the macroscopic models that we have obtained from the kinetic description of the particle models.

In all tests we consider the strongly connected graph illustrated in Figure~\ref{fig:network3nodes}, which models a network of e.g., three different cities. Hence $\cI=\{1,\,2,\,3\}$. We choose the transition matrix as
\begin{equation}
	\bP=
		\begin{pmatrix}
			P_{11} & P_{12} & P_{13} \\
			P_{21} & P_{22} & P_{23} \\
			P_{31} & P_{32} & P_{33}
		\end{pmatrix}
	=
		\begin{pmatrix}
			0.3 & 0.1 & 0 \\
			0.7 & 0.7 & 0.5 \\
			0 & 0.2 & 0.5
		\end{pmatrix}
	\label{eq:P.num}
\end{equation}
and the parameters of the interactions as indicated in Table~\ref{tab:param}.

\subsection{Test 1: Basic interaction dynamics on a network}
\label{sect:num.test_1}
\begin{figure}[!t]
\centering
\subfigure[]{\includegraphics[width=.47\textwidth]{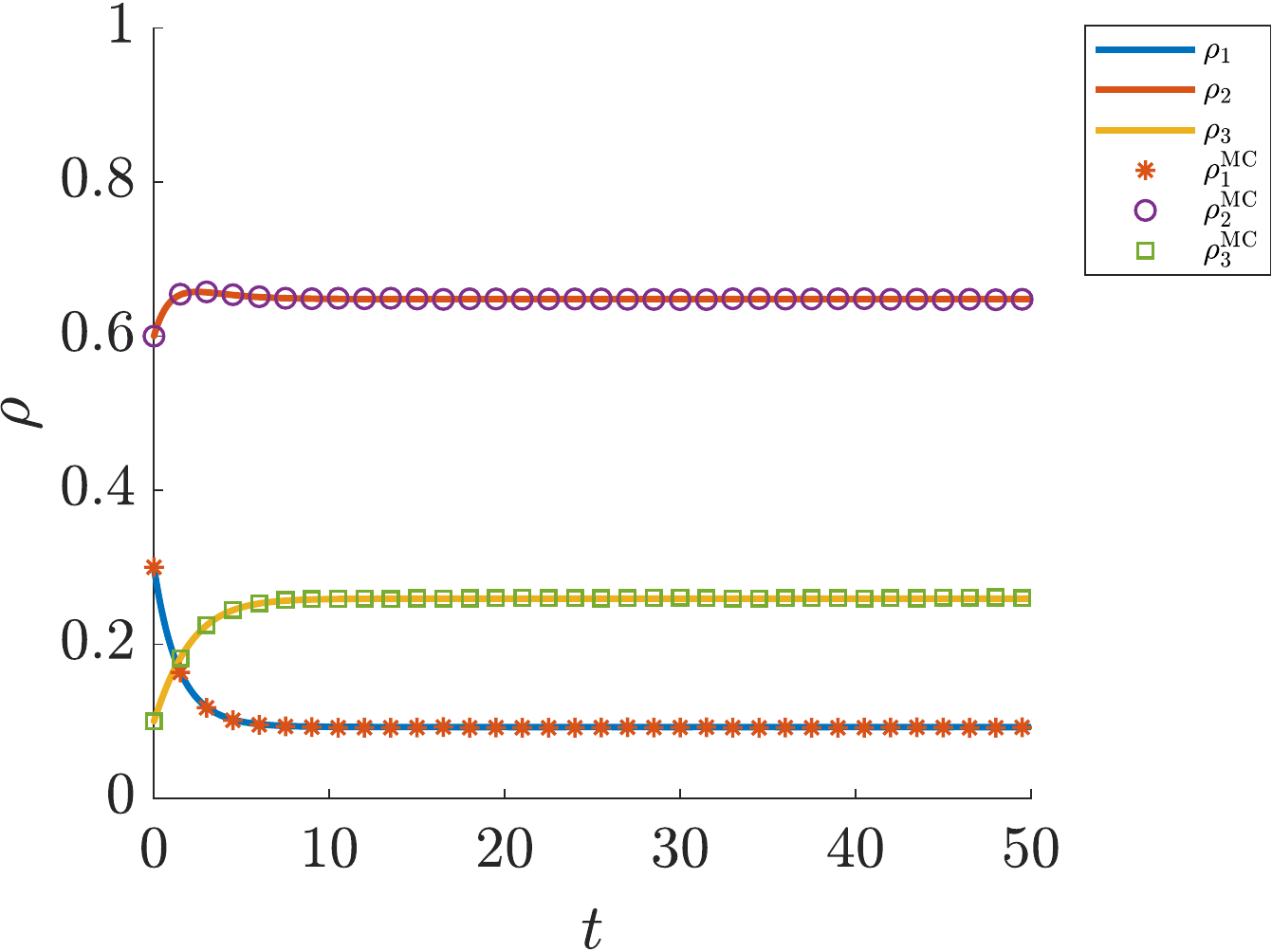}} \qquad
\subfigure[]{\includegraphics[width=.47\textwidth]{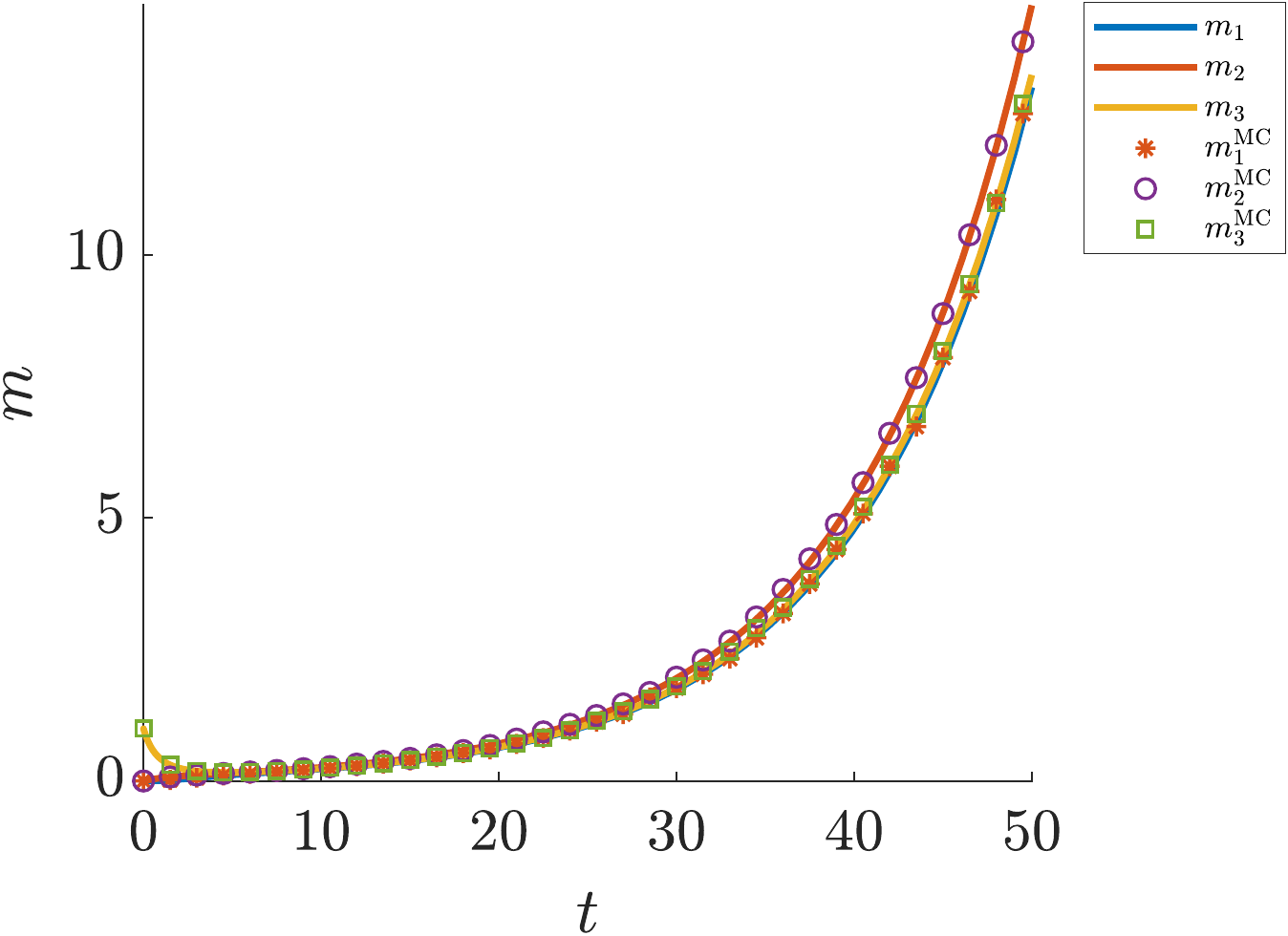}} \\
\subfigure[]{\includegraphics[width=.47\textwidth]{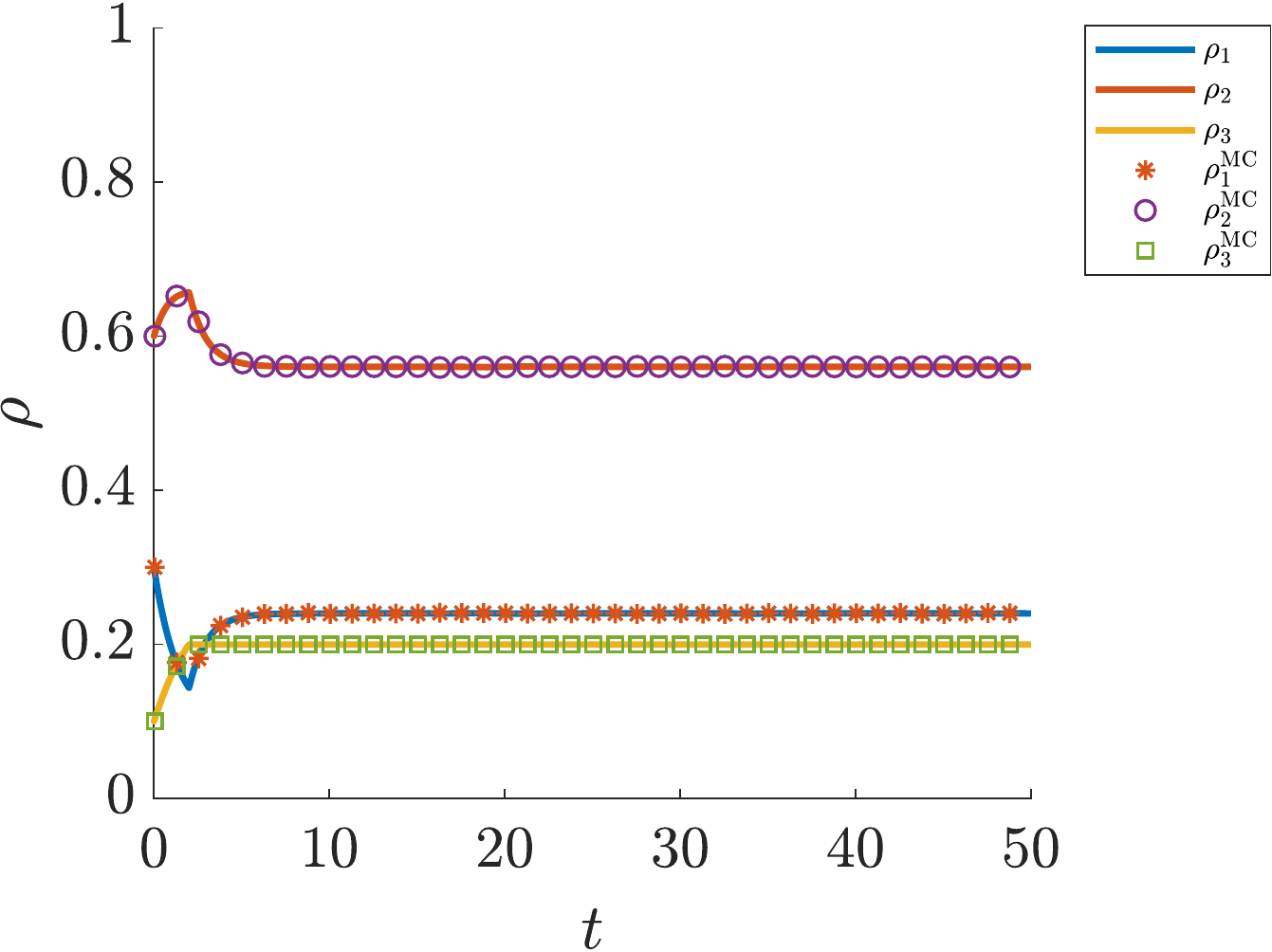}} \qquad
\subfigure[]{\includegraphics[width=.47\textwidth]{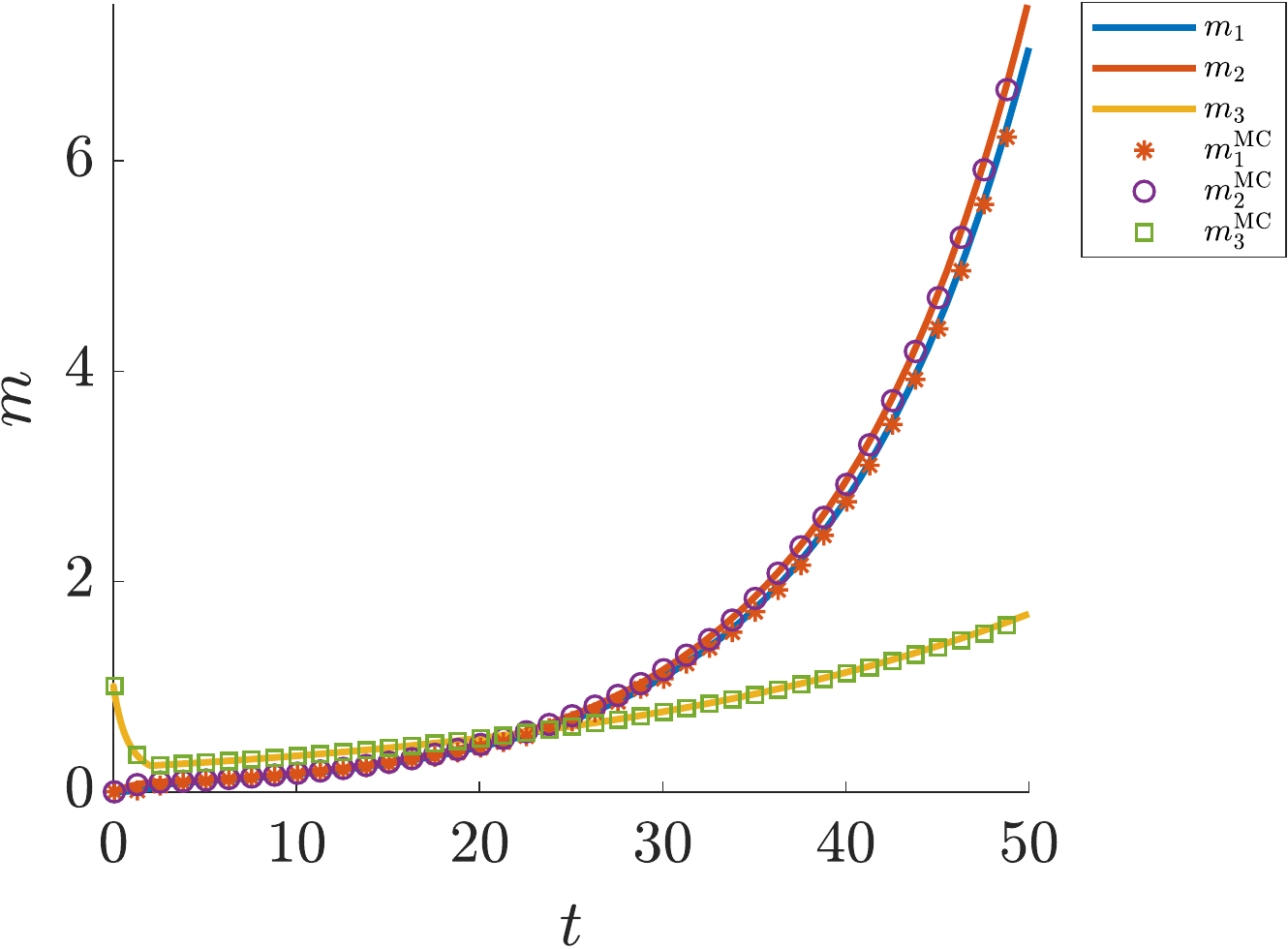}}
\caption{Test 1 (basic interaction dynamics on a network, Section~\ref{sect:num.test_1}). Top row: standard evolution. Bottom row: \textit{cordon sanitaire} around node $3$ for $t\geq 2$}
\label{fig:net_int}
\end{figure}

We begin from the basic model described in Section~\ref{sect:Boltzmann.graph}, in which individuals move across the nodes of the network according to the transition probabilities encoded in the matrix~\eqref{eq:P.num} and exchange viral loads according to the interaction rules~\eqref{eq:binary_gen_sym}-\eqref{eq:pq}. The particle model is~\eqref{eq:particle.gen} while the corresponding macroscopic model is~\eqref{eq:rho_i}-\eqref{eq:m_i}.

We choose the initial conditions as follows: in nodes $1$ and $2$ we set
$$ f_{1,0}(v)=0.3\delta(v), \qquad f_{2,0}(v)=0.6\delta(v) $$
while in node $3$ we let $f_{3,0}(v)$ be an inverse-gamma distribution with shape parameter $3$, scale parameter $2$ and mass equal to $0.1$:
$$ f_{3,0}(v)=\frac{2e^{-\frac{2}{v}}}{5v^4} $$
Therefore initially nodes $1$ and $2$ are disease-free, i.e. $m_{1,0}=m_{2,0}=0$, while node $3$ features an infection onset with $m_{3,0}=1$. The initial densities are $\rho_{1,0}=0.3$, $\rho_{2,0}=0.6$, $\rho_{3,0}=0.1$.

Figures~\ref{fig:net_int}a,~\ref{fig:net_int}b show that the densities in the three nodes reach non-zero asymptotic values as predicted by Proposition~\ref{prop:rho^inf} and that, since $\nu_1<\nu_2$, the infection blows up in all nodes consistently with the analysis performed in Section~\ref{sect:aggregate}. These figures also show a perfect correspondence between the Monte Carlo solution of the particle model~\eqref{eq:particle.gen} and the macroscopic model~\eqref{eq:rho_i}-\eqref{eq:m_i}.

Figures~\ref{fig:net_int}c,~\ref{fig:net_int}d show instead the effect of a \textit{cordon sanitaire} around node $3$, which from time $t=2$ onwards isolates that node from the others. We simulate this by modifying the transition matrix~\eqref{eq:P.num} for $t\geq 2$ as
$$ \bP=
	\begin{pmatrix}
		0.3 & 0.3 & 0 \\
		0.7 & 0.7 & 0 \\
		0 & 0 & 1
	\end{pmatrix}. $$
In particular, $P_{33}=1$ and $P_{i3}=P_{3j}=0$ for $i,\,j=1,\,2$ because the \textit{cordon sanitaire} prevents individuals from entering and exiting from node $3$. The densities reach new non-zero asymptotic values and the infection still blows up in all nodes because $\nu_1<\nu_2$. Hence isolating the hotspot of the infection is \textit{per se} useless as a confinement measure if it is not associated to other targeted interventions within the nodes. We will come back to this issue in Section~\ref{sect:num.test_3}. For the moment, we observe however that the model predicts a lower rate of blow-up of the infection in all nodes compared to the case without \textit{cordon sanitaire}. In particular, despite the fact that the infection originated from node $3$, the latter features the lowest blow-up rate, because it has been isolated from node $2$, which is the most populated and connected one. Again, we notice a nice matching between the solutions of the particle and the macroscopic models.

\subsection{Test 2: Commuters}
\label{sect:num.test_2}
\begin{figure}[!t]
\centering
\includegraphics[width=.55\textwidth]{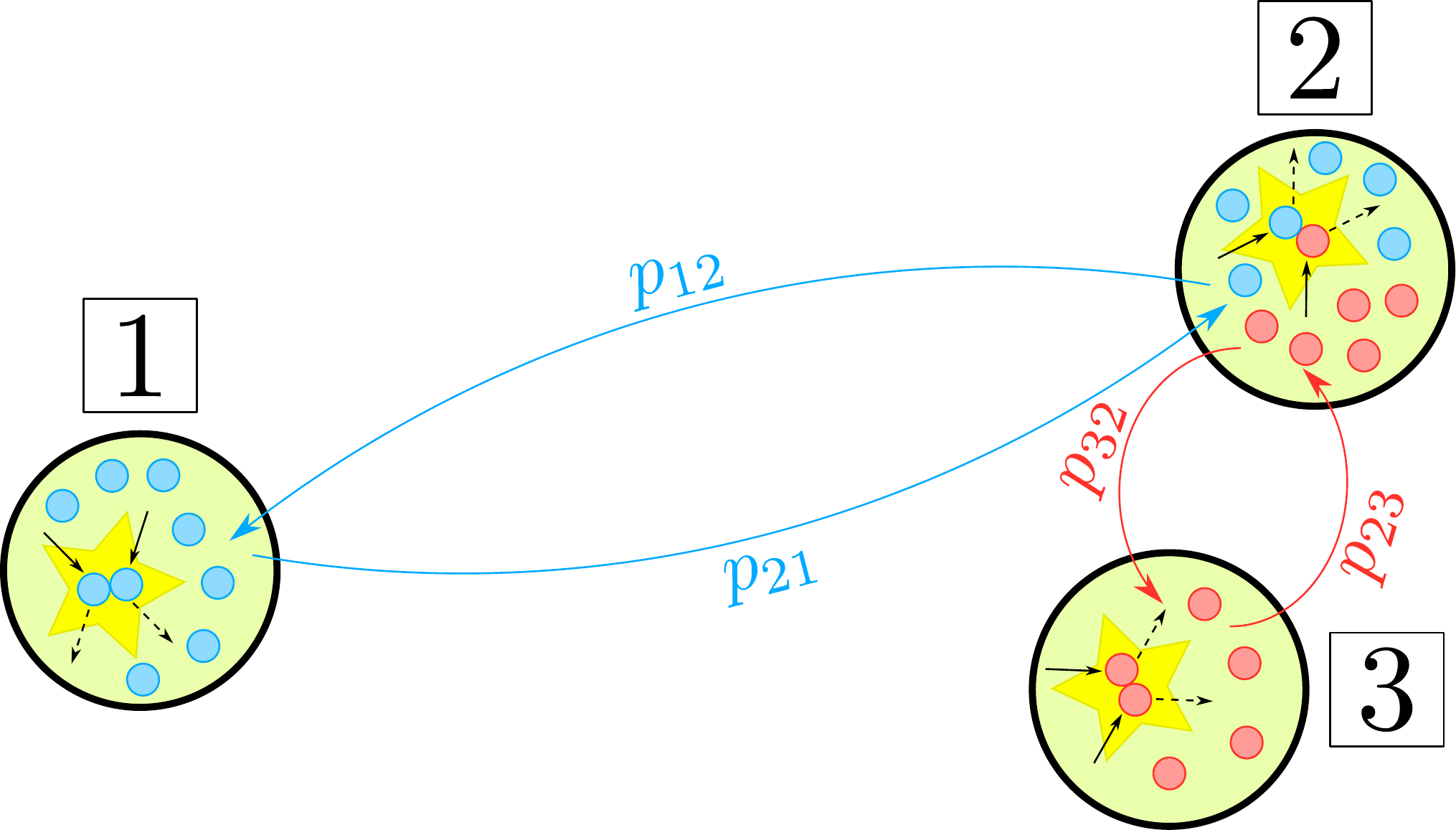}
\caption{The three-node network for the problem of the commuters discussed in Section~\ref{sect:num.test_2}. Blue particles travel only between nodes $1$ and $2$ while red particles travel only between nodes $2$ and $3$. Moreover, in every node there is a portion of particles that do not commute}
\label{fig:network3nodes_commuters}
\end{figure}

We consider now the model of the commuters introduced in Section~\ref{sect:commuters}, cf. also Figure~\ref{fig:network3nodes_commuters}. We choose the following commuting probabilities:
\begin{equation*}
	\begin{array}{ll}
		p_{21}=0.7, & p_{31}=0, \\
		p_{12}=0.3, & p_{32}=0.8, \\
		p_{13}=0, & p_{23}=0.2,
	\end{array}
\end{equation*}
and we set $p_{ii}=1$ for all $i=1,\,2,\,3$. We recall that, as explained in Remark~\ref{rem:commuters}, the conceptual meaning of these probabilities is different from that of the transition probabilities~\eqref{eq:P.num}. In particular, the latter are not used in this model.

At $t=0$ we prescribe the following distributions:
\begin{equation*}
	\begin{array}{lll}
		f_{11,0}(v)=0.21\delta(v), & f_{12,0}(v)=0.06\delta(v), & f_{13,0}(v)=0, \\
		f_{21,0}(v)=0.09\delta(v), & f_{22,0}(v)=0.3\delta(v), & f_{23,0}(v)=0.05e^{-v}, \\
		f_{31,0}(v)=0, & f_{32,0}(v)=0.24\delta(v), & f_{33,0}(v)=0.05e^{-v},
	\end{array}
\end{equation*}
meaning that initially all commuting routes are disease-free but the routes $3\to 2$ and $3\to 3$, where $m_{32,0}=m_{33,0}=1$. Therefore node $3$ is again the infection hotspot. As a matter of fact, $3\to 3$ is actually not a real commuting route but identifies individuals who remain always in node $3$, i.e. do not commute.

The total masses of commuters initially present in the nodes are
$$ \rho_{1,0}=0.3, \quad \rho_{2,0}=0.6, \quad \rho_{3,0}=0.1, $$
while those of the commuters along the various routes are
$$ \bar{\rho}_{\{1,2\}}=0.15, \quad \bar{\rho}_{\{1,3\}}=0, \quad \bar{\rho}_{\{2,3\}}=0.29. $$
As explained in Section~\ref{sect:commuters}, these latter values are conserved in time. We also compute
$$ \bar{\rho}_{\{1,1\}}=0.42, \quad \bar{\rho}_{\{2,2\}}=0.6, \quad \bar{\rho}_{\{3,3\}}=0.1, $$
noticing that $\bar{\rho}_{\{i,\,i\}}=2\rho_{ii,0}$ for all $i=1,\,2,\,3$ for consistency with the definition of $\bar{\rho}_{\{i,\,j\}}$ given in Section~\ref{sect:commuters}.

The particle model is now~\eqref{eq:particle.commuters} and the macroscopic model is~\eqref{eq:rhoji}-\eqref{eq:mji}.

\begin{figure}[!t]
\centering
\subfigure[]{\includegraphics[width=.47\textwidth]{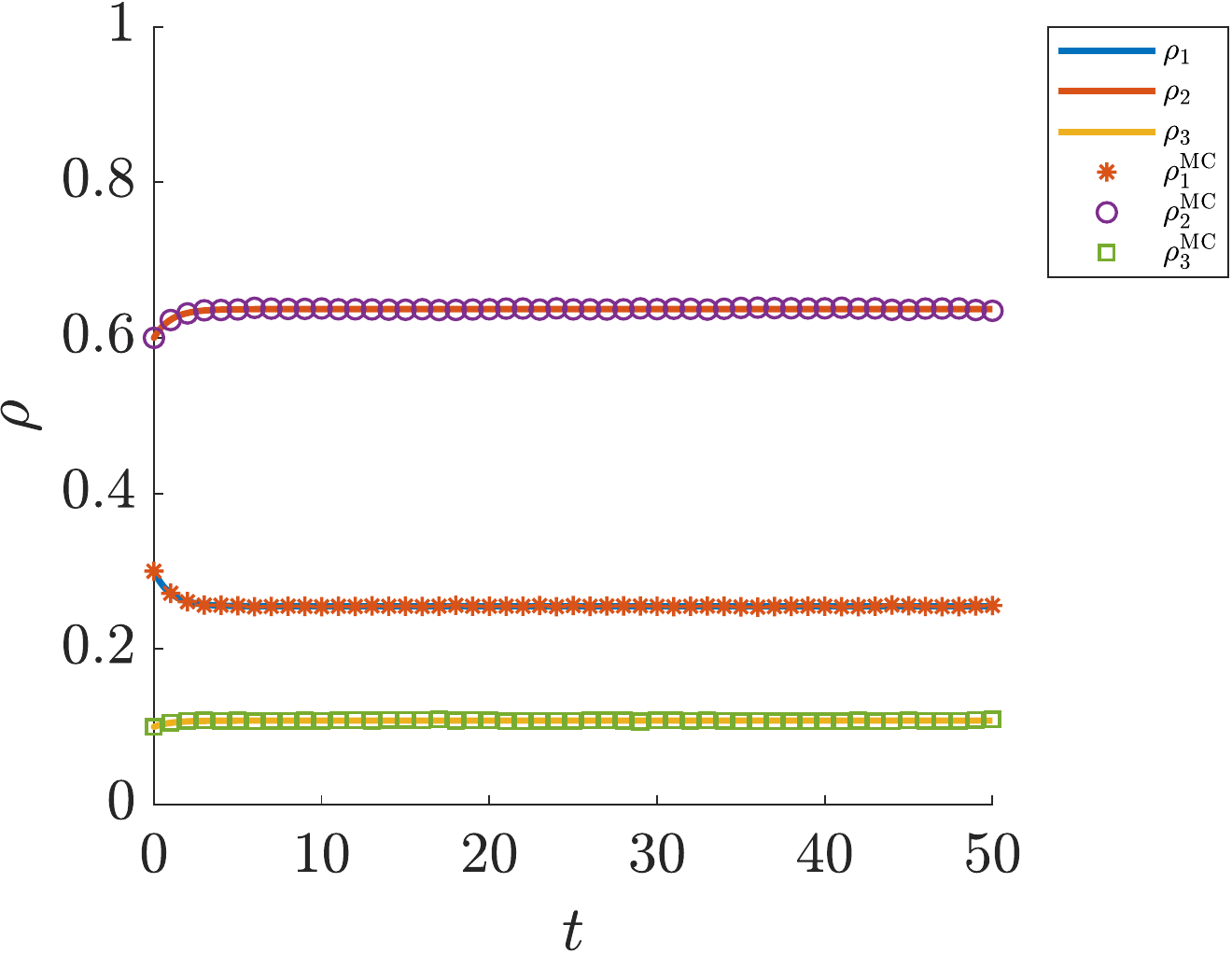}} \qquad
\subfigure[]{\includegraphics[width=.47\textwidth]{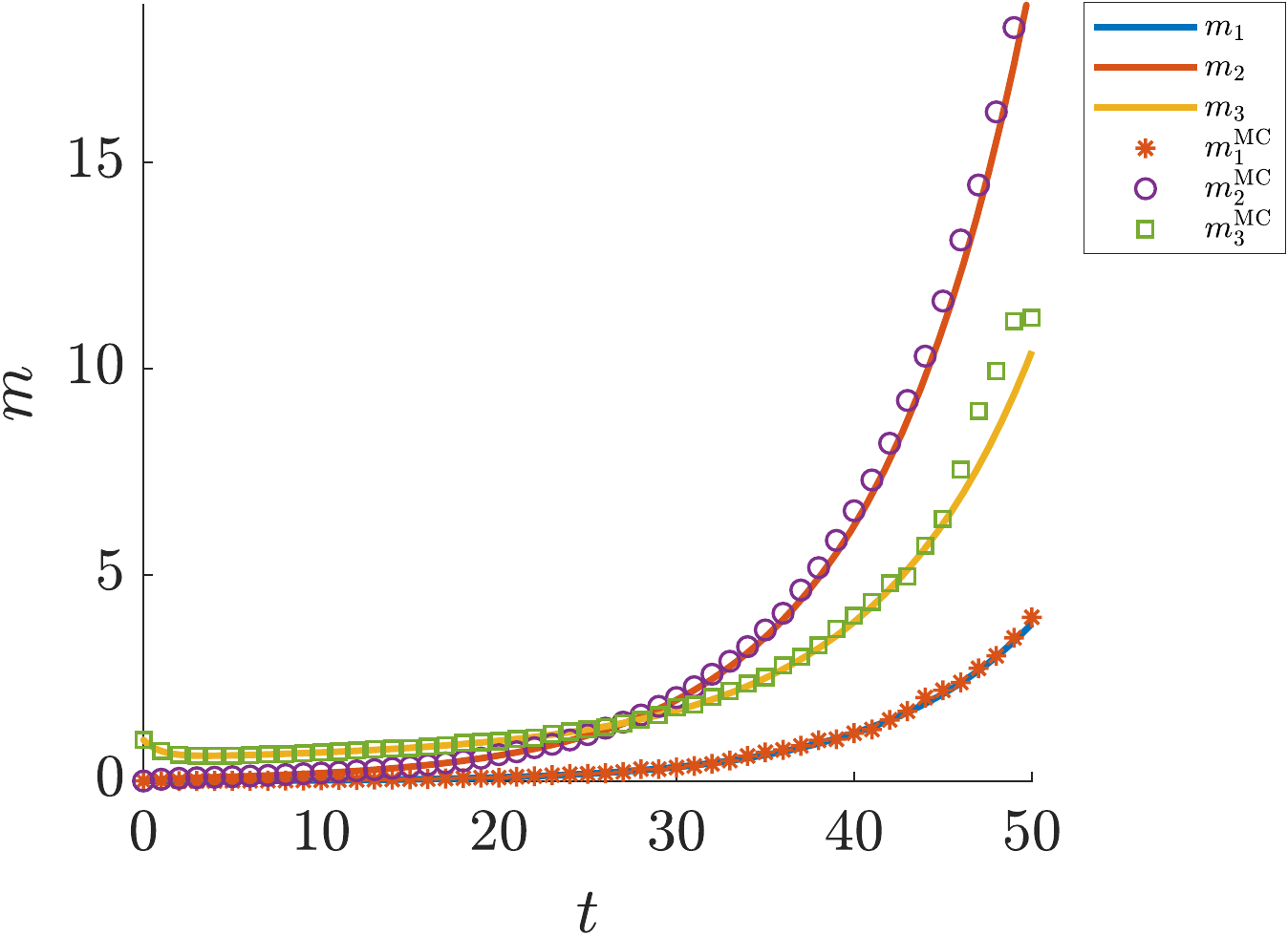}}
\caption{Test 2 (commuters, Section~\ref{sect:num.test_2})}
\label{fig:commuters}
\end{figure}

From Figure~\ref{fig:commuters}a we see that the total densities in the nodes quickly reach non-zero asymptotic values, which, plugging the numerical data above into~\eqref{eq:rho^inf_i}, can be computed precisely as
$$ \rho_1^\infty=0.255, \quad \rho_2^\infty=0.637, \quad \rho_3^\infty=0.108. $$
We remark that, due to the repetitiveness of the commuting dynamics, it makes perfectly sense that the densities in the nodes are essentially constant in time apart from a short initial transient. We observe moreover that the gap $\abs{\rho_i^\infty-\rho_{i,0}}$ between the initial and equilibrium values is of order $10^{-2}$, hence reasonably small, for all $i=1,\,2,\,3$.

From Figure~\ref{fig:commuters}b we observe once again a blow-up of the infection in all nodes like in Test 1. Nevertheless, unlike the case illustrated in Figure~\ref{fig:net_int}b, here the trends of the mean viral loads in the three nodes differ more consistently from each other. In particular, the blow-up occurs earlier in node $2$, i.e. the one more crossed by commuting routes, and later in node $1$, which is not only one of the two nodes less crossed by commuting routes but also an initially disease-free node. This shows that if one assumes more specific migration paths than simple random transitions from node to node then our models can provide more accurate and realistic predictions.

Finally, from Figure~\ref{fig:commuters} we still observe a nice agreement between the microscopic particle dynamics and the aggregate macroscopic trends derived through the kinetic description.

\subsection{Test 3: Quarantine in the nodes}
\label{sect:num.test_3}
Finally we consider the model introduced in Section~\ref{sect:quarantine} to explore the effect of quarantine as a measure to confine the infection node-wise. Remarkably, quarantine is one of the few confinement measures immediately implementable in case of new infectious diseases for which medical treatments are not yet available.

The network for this test is the one in Figure~\ref{fig:net_int} with transition probabilities $P^1_{ij}$ of non-quarantined individuals coinciding with the corresponding $P_{ij}$'s in~\eqref{eq:P.num}. The transition probabilities of quarantined individuals are instead the trivial ones $P^2_{ij}=\delta_{ij}$, consistently with the general setting presented in Section~\ref{sect:quarantine}. The other relevant parameters of the interactions are given in Table~\ref{tab:param}. In addition to them, we need to prescribe the label switch probabilities $T^{21}_i(v)$, $T^{12}_i(v)$, which describe the processes by which an individual with viral load $v$ in node $i$ is diagnosed as infected, hence put in isolation, and released from quarantine, hence readmitted in the society, respectively. We observe that $T^{21}_i(v)$, $T^{12}_i(v)$ may be correlated with the sensitivity of either testing technique used to detect the infection in an individual.

The initial conditions are the following:
\begin{align*}
	& f^1_{1,0}(v)=0.3\delta(v), \quad f^1_{2,0}(v)=0.6\delta(v), \quad f^1_{3,0}(v)=0.2e^{-2v}, \\
	& f^2_{i,0}(v)=0, \quad \forall\,i=1,\,2,\,3.
\end{align*}
They model a scenario in which at $t=0$ no individual is quarantined in any node and node $3$ is the hotspot of the infection. In particular, we have
$$ \rho_{1,0}=\rho^1_{1,0}=0.3, \quad \rho_{2,0}=\rho^1_{2,0}=0.6, \quad \rho_{3,0}=\rho^1_{3,0}=0.1 $$
and
$$ \rho^2_{1,0}=\rho^2_{2,0}=\rho^2_{3,0}=0. $$
Furthermore, $m_{3,0}=m^1_{3,0}=0.5$ while all other mean viral loads are initially zero.

We begin with the case of constant label switch probabilities: then the particle model~\eqref{eq:particle.quarantine} admits a macroscopic counterpart in closed form given by model~\eqref{eq:rhoih.system}-\eqref{eq:mih.system}. In particular, we let
$$ T^{21}_i=0.2, \quad T^{12}_i=0.4, \qquad \forall\,i\in\{1,\,2,\,3\}. $$

\begin{figure}[!t]
\centering
\subfigure[]{\includegraphics[width=.47\textwidth]{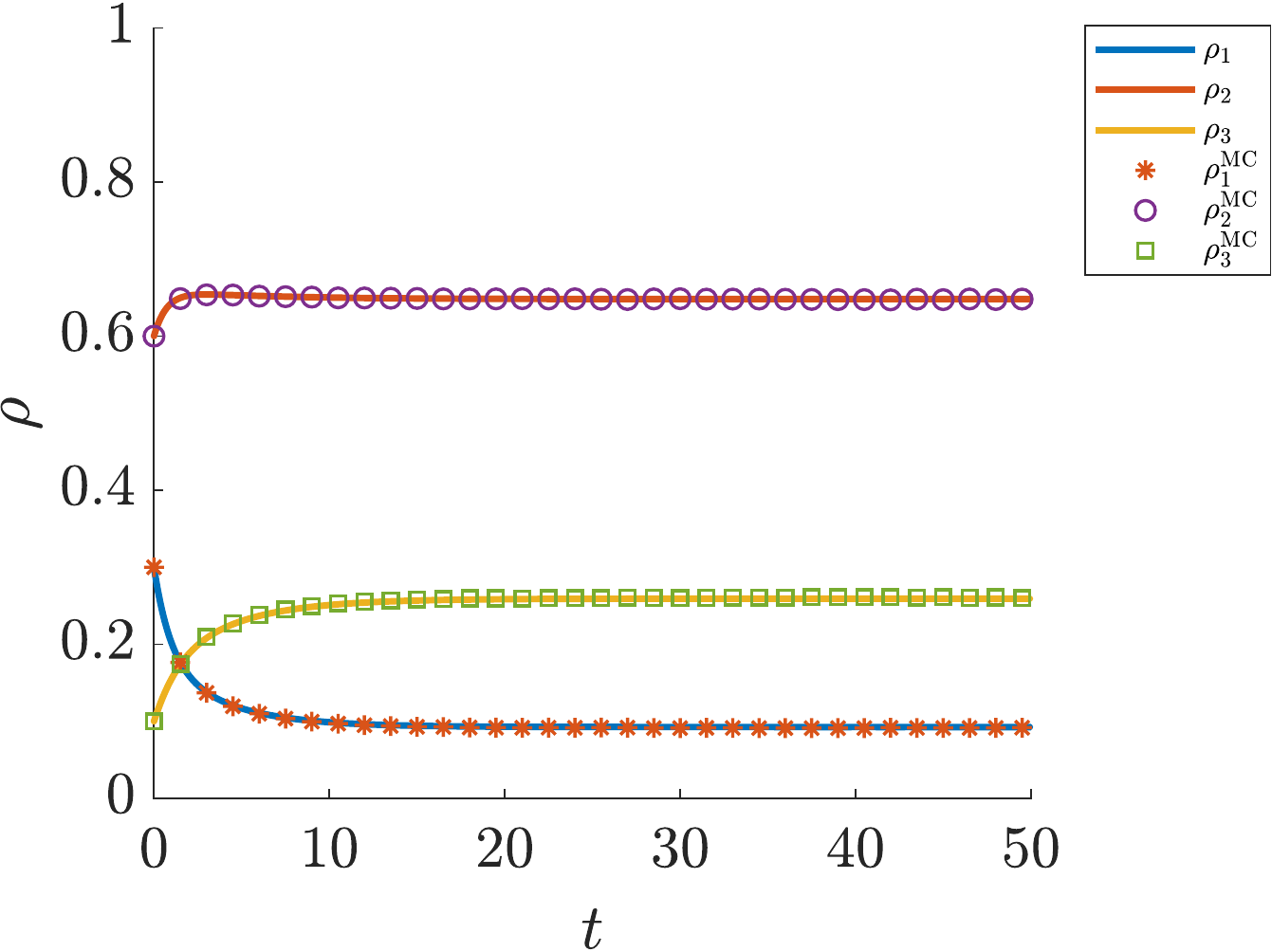}} \qquad
\subfigure[]{\includegraphics[width=.47\textwidth]{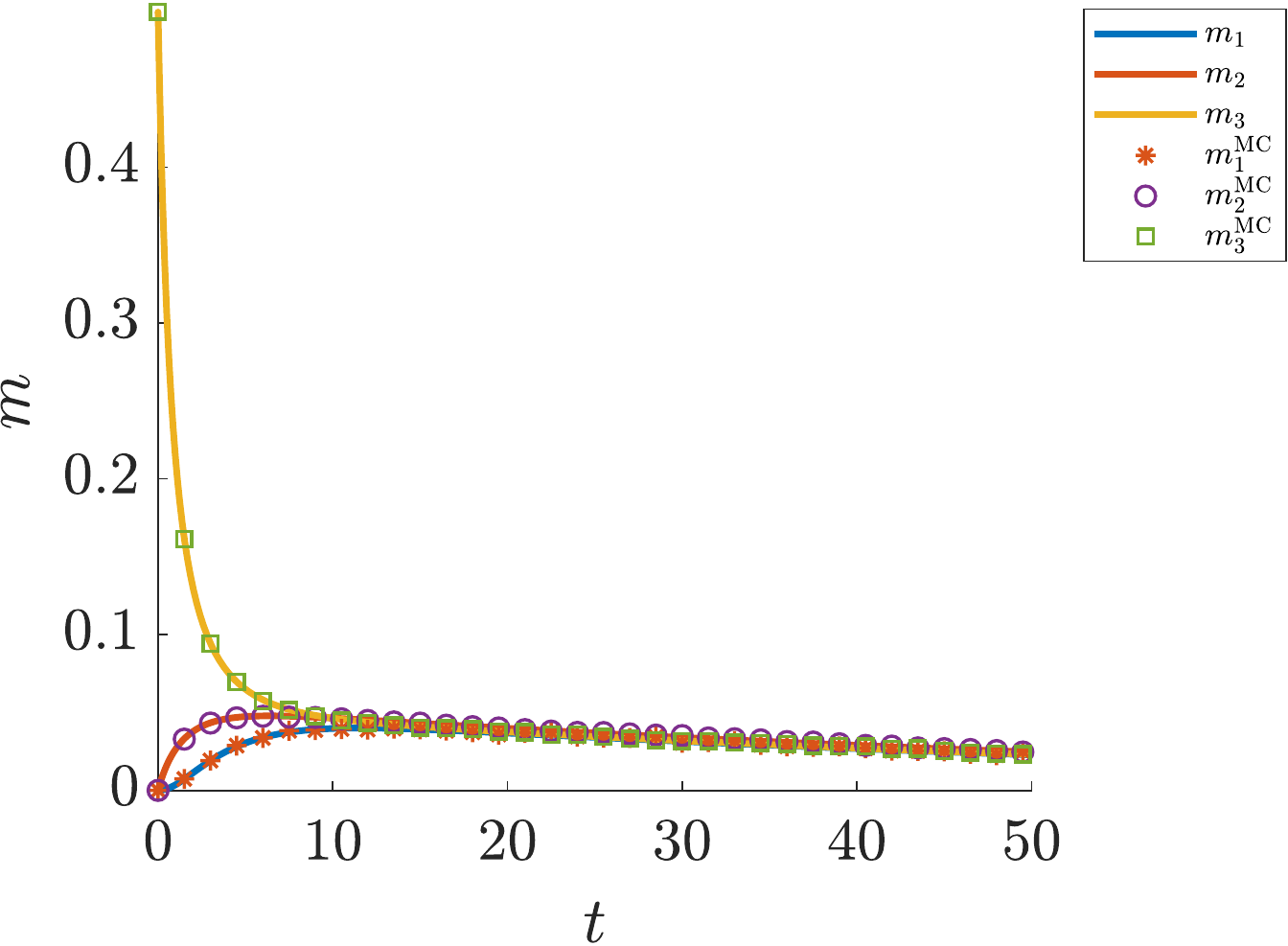}} \\
\subfigure[]{\includegraphics[width=.47\textwidth]{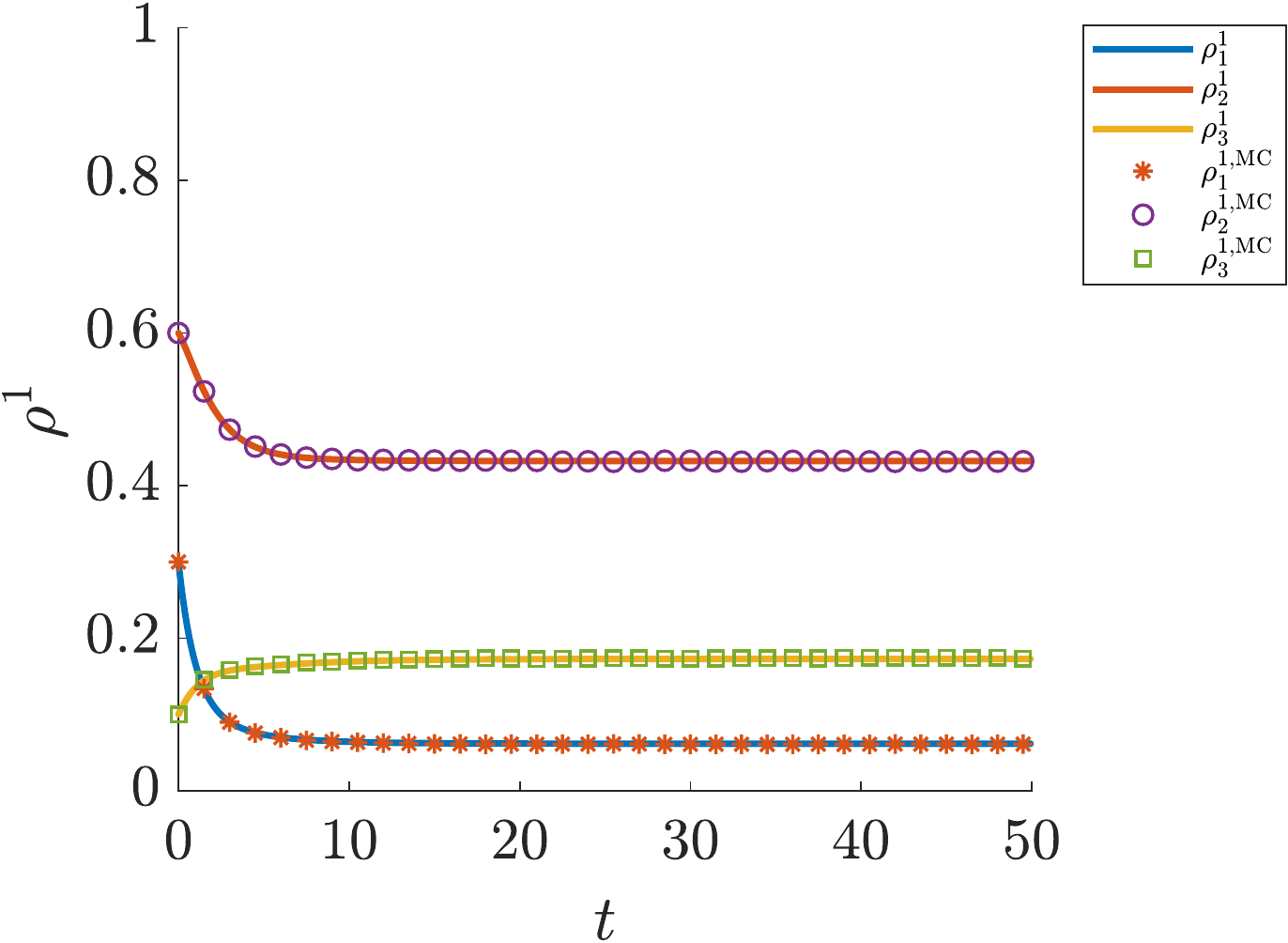}} \qquad
\subfigure[]{\includegraphics[width=.47\textwidth]{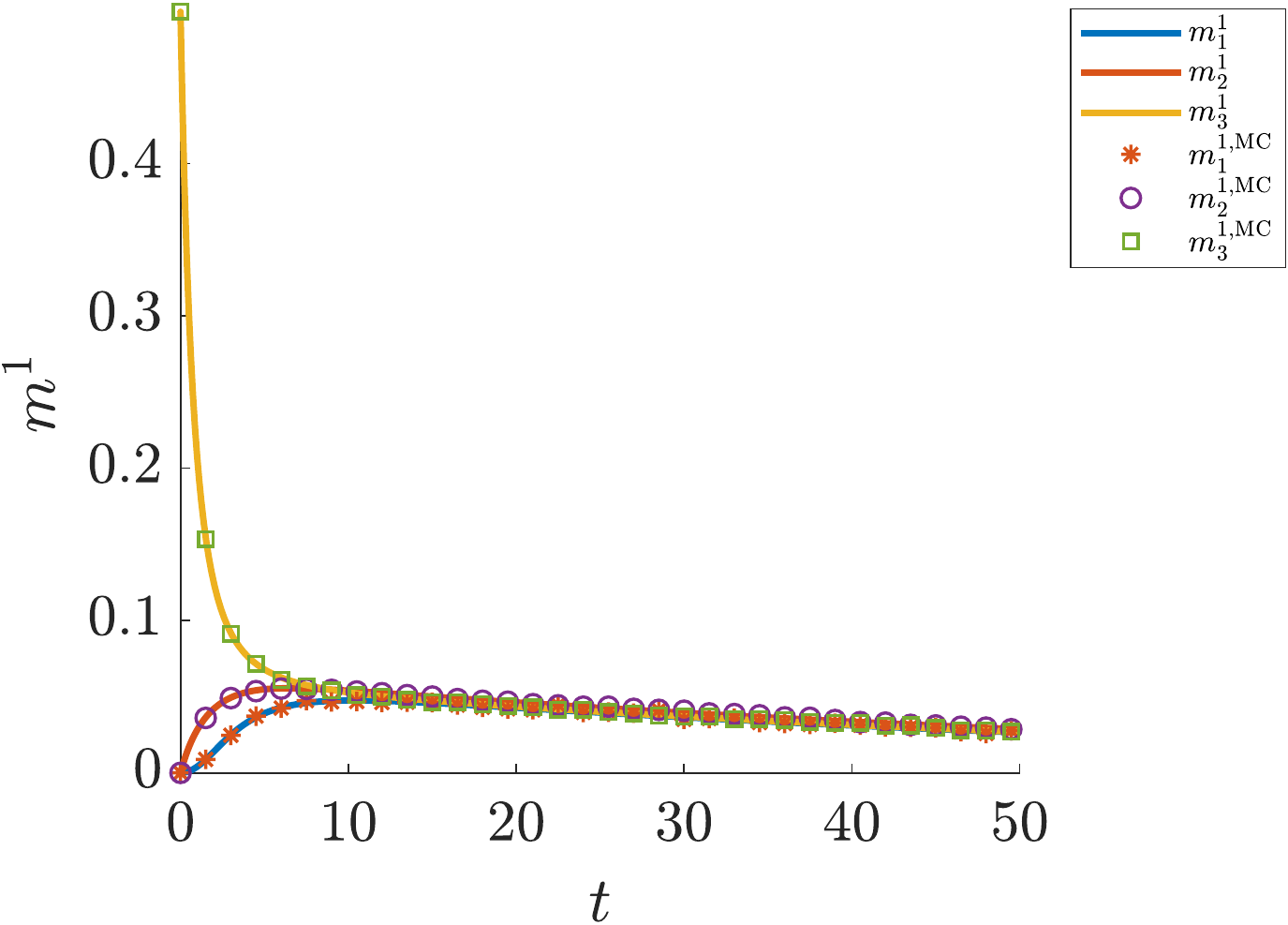}} \\
\subfigure[]{\includegraphics[width=.47\textwidth]{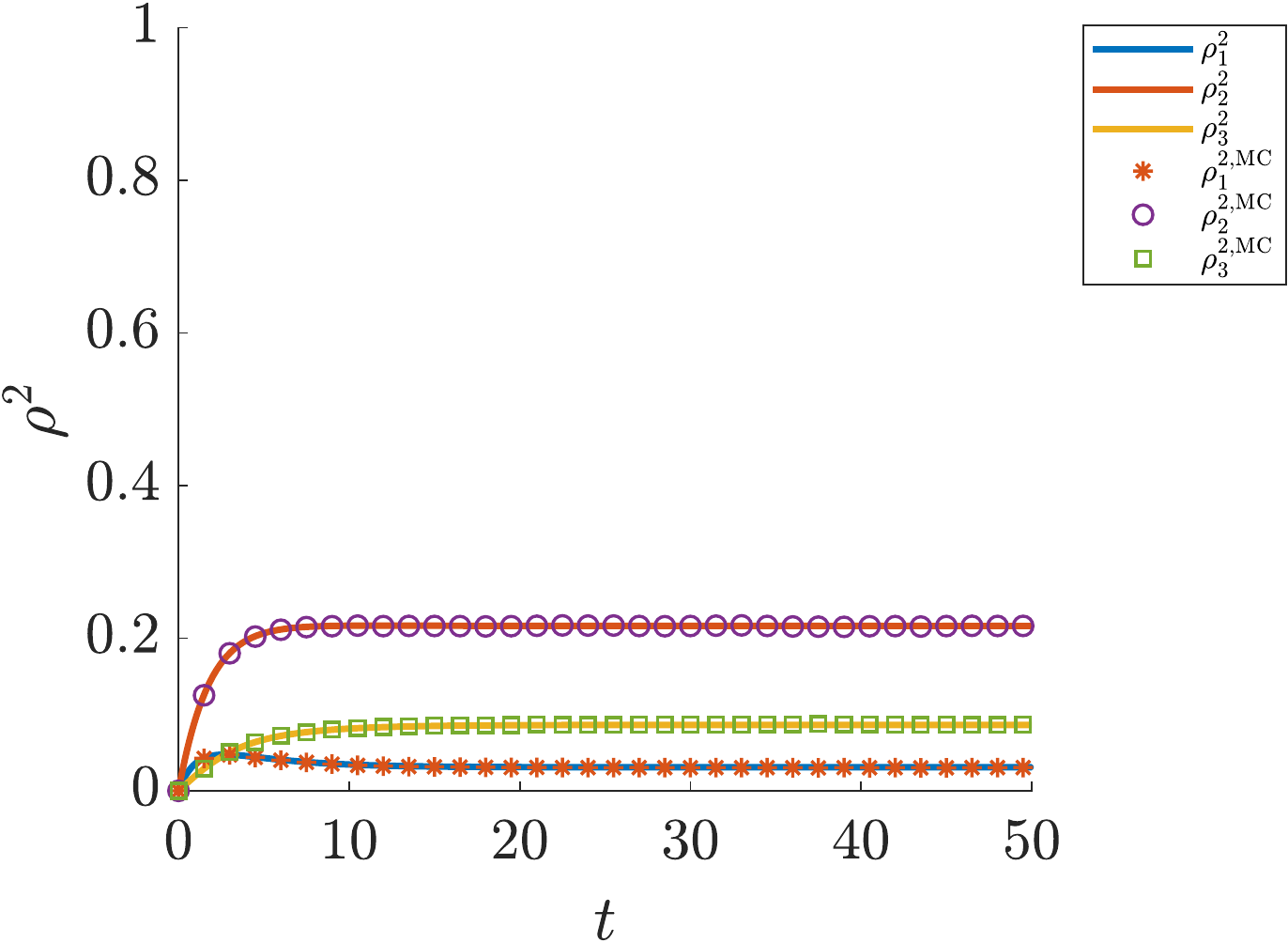}} \qquad
\subfigure[]{\includegraphics[width=.47\textwidth]{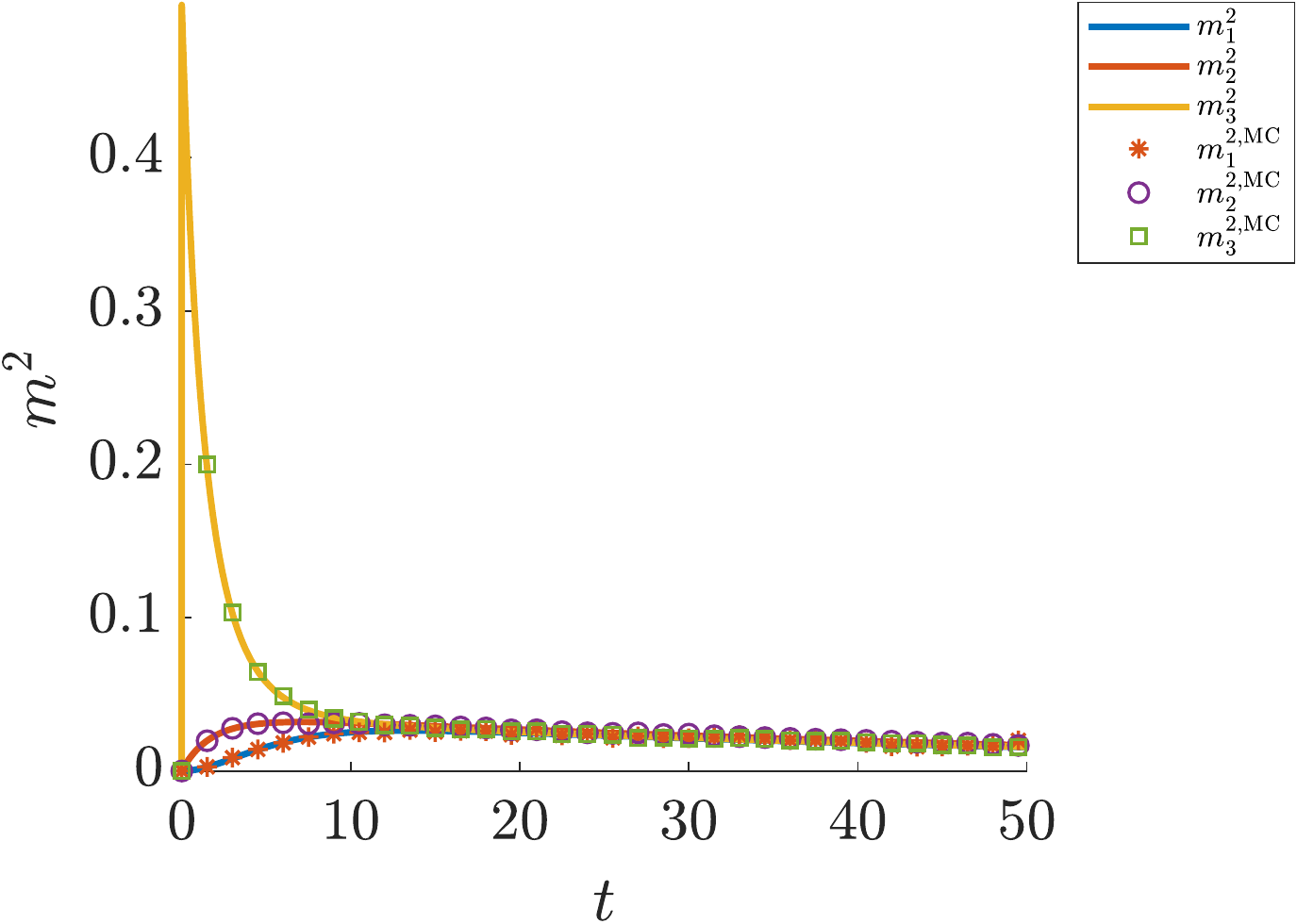}} \\
\caption{Test 3 (quarantine in the nodes, Section~\ref{sect:num.test_3}) with constant label switch probabilities}
\label{fig:quarantine-cp0}
\end{figure}

Figure~\ref{fig:quarantine-cp0} shows the time trends of the total density and mean viral load (top row), as well as those of the density and mean viral load of non-quarantined and quarantined individuals (middle and bottom rows, respectively), in each node of the network. Apart from remarking again a perfect correspondence between the solutions to the particle and the macroscopic models, we observe from Figure~\ref{fig:quarantine-cp0}b (and, with further specificity, from Figures~\ref{fig:quarantine-cp0}d,~\ref{fig:quarantine-cp0}f) that in the long run the infection is eradicated in every node because all mean viral loads tend to zero for $t\to +\infty$. It is interesting to compare this result with the one obtained in~\cite{loy2021KRM_preprint}, where the same model with quarantine is analysed in the absence of a network. In practice, the model in~\cite{loy2021KRM_preprint} can be seen as the analogous of the present model in an isolated node, for instance one around which a \textit{cordon sanitaire} has been established. Remarkably, in that case and with the very same values of the parameters used here an infection blow-up occurs (cf.~\cite[Figure~2]{loy2021KRM_preprint}). Only if the probability for an individual to be diagnosed as infected and quarantined is sufficient higher than $0.2$ (the analysis in~\cite[Sections~5.1,~6]{loy2021KRM_preprint} indicates that the minimal threshold is $0.28$) the infection may be successfully kept under control and eradicated in the long run. The comparison between these two results puts in evidence a striking effect of the network: allowing the individuals to migrate and mix in different locations, with all other features unchanged including the effectiveness of the testing techniques, may help dominate the infection more effectively than the confinement produced by a \textit{cordon sanitaire}. This conclusion, here emerging naturally as a consequence of our quite general model, is in agreement with the observations made in~\cite{espinoza2020PLOS1} using more \textit{ad-hoc} compartmental models specifically conceived to address this phenomenon.

\begin{figure}[!t]
\centering
\subfigure[]{\includegraphics[width=.47\textwidth]{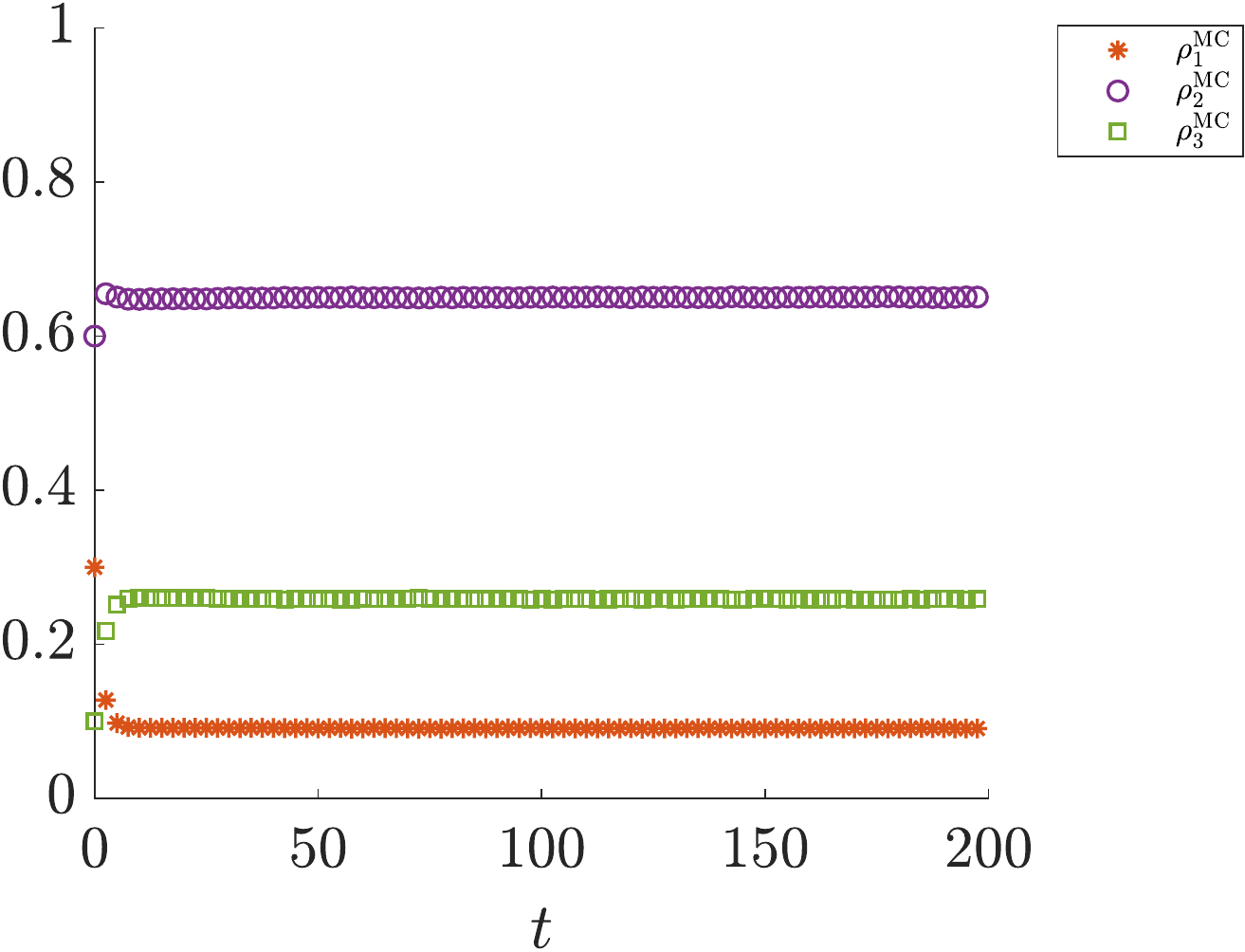}} \qquad
\subfigure[]{\includegraphics[width=.47\textwidth]{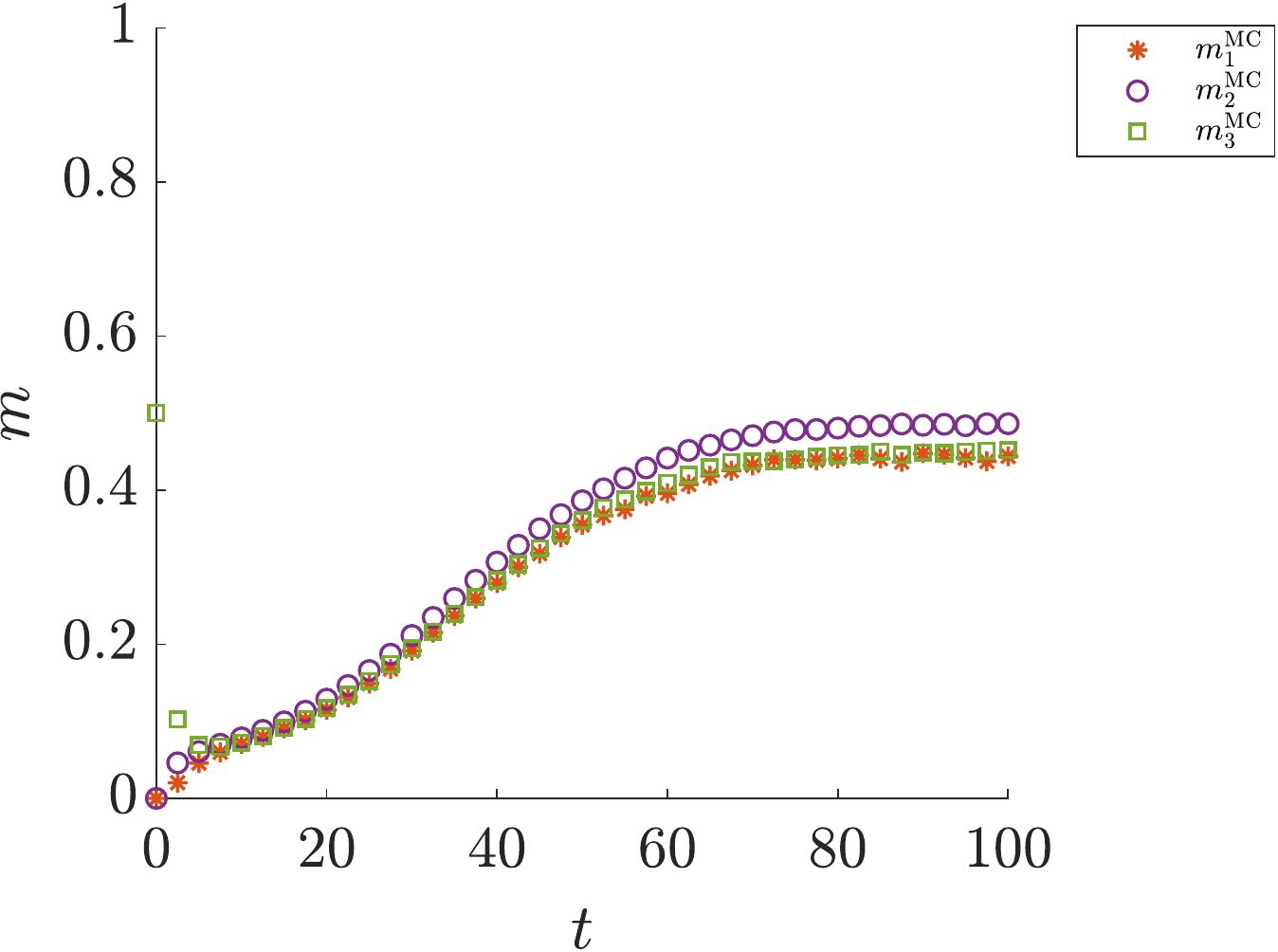}} \\
\subfigure[]{\includegraphics[width=.47\textwidth]{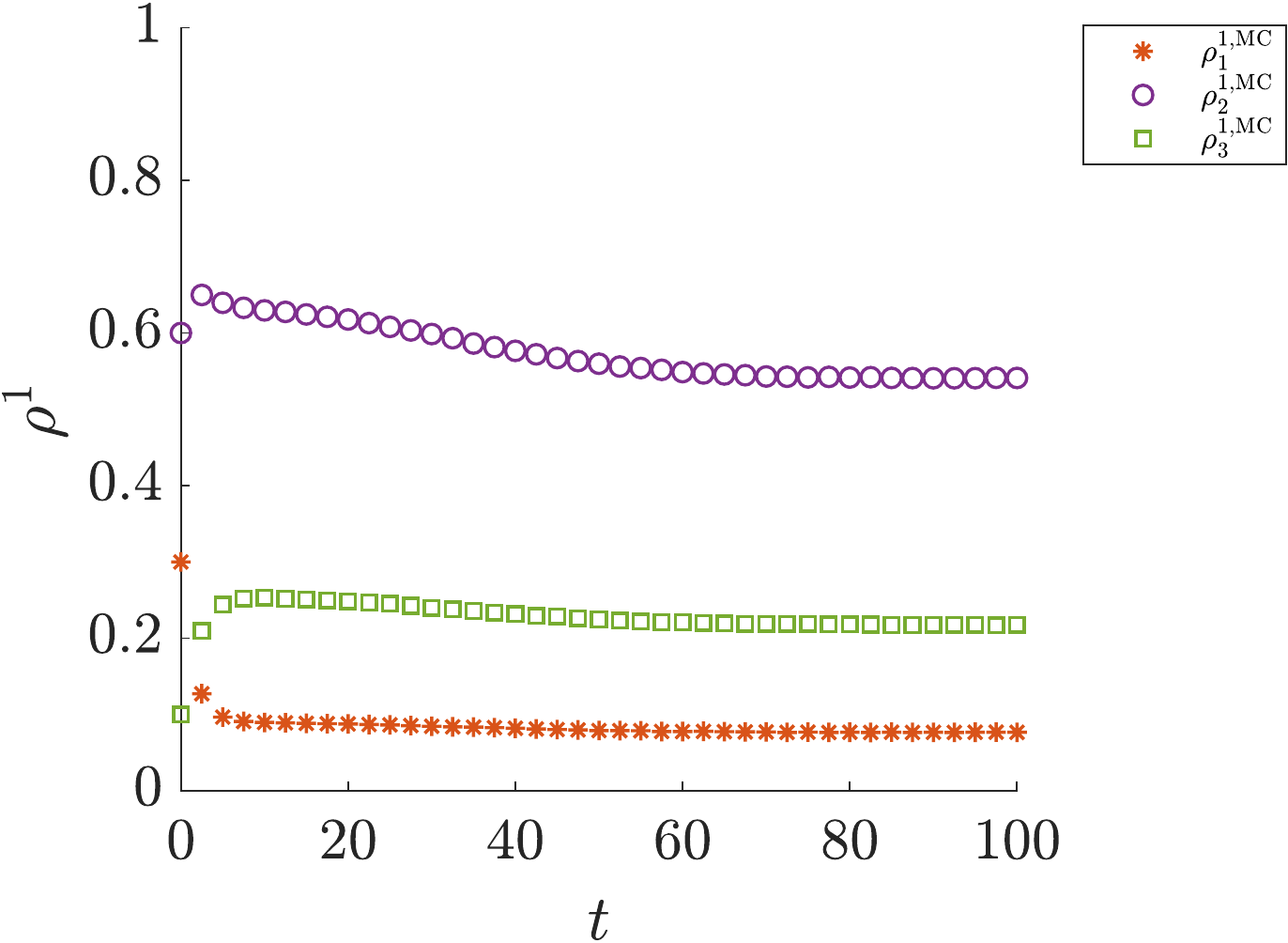}} \qquad
\subfigure[]{\includegraphics[width=.47\textwidth]{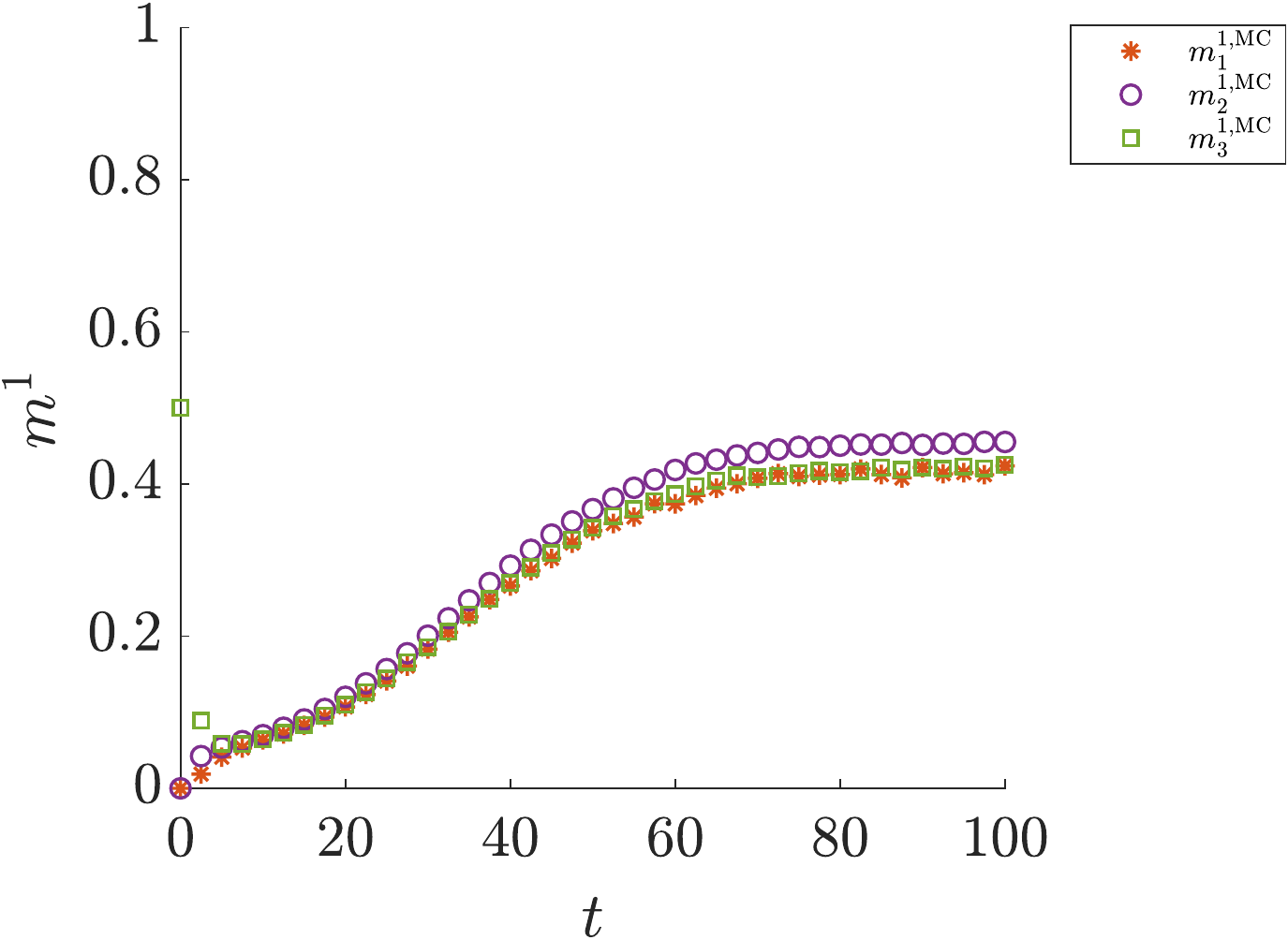}} \\
\subfigure[]{\includegraphics[width=.47\textwidth]{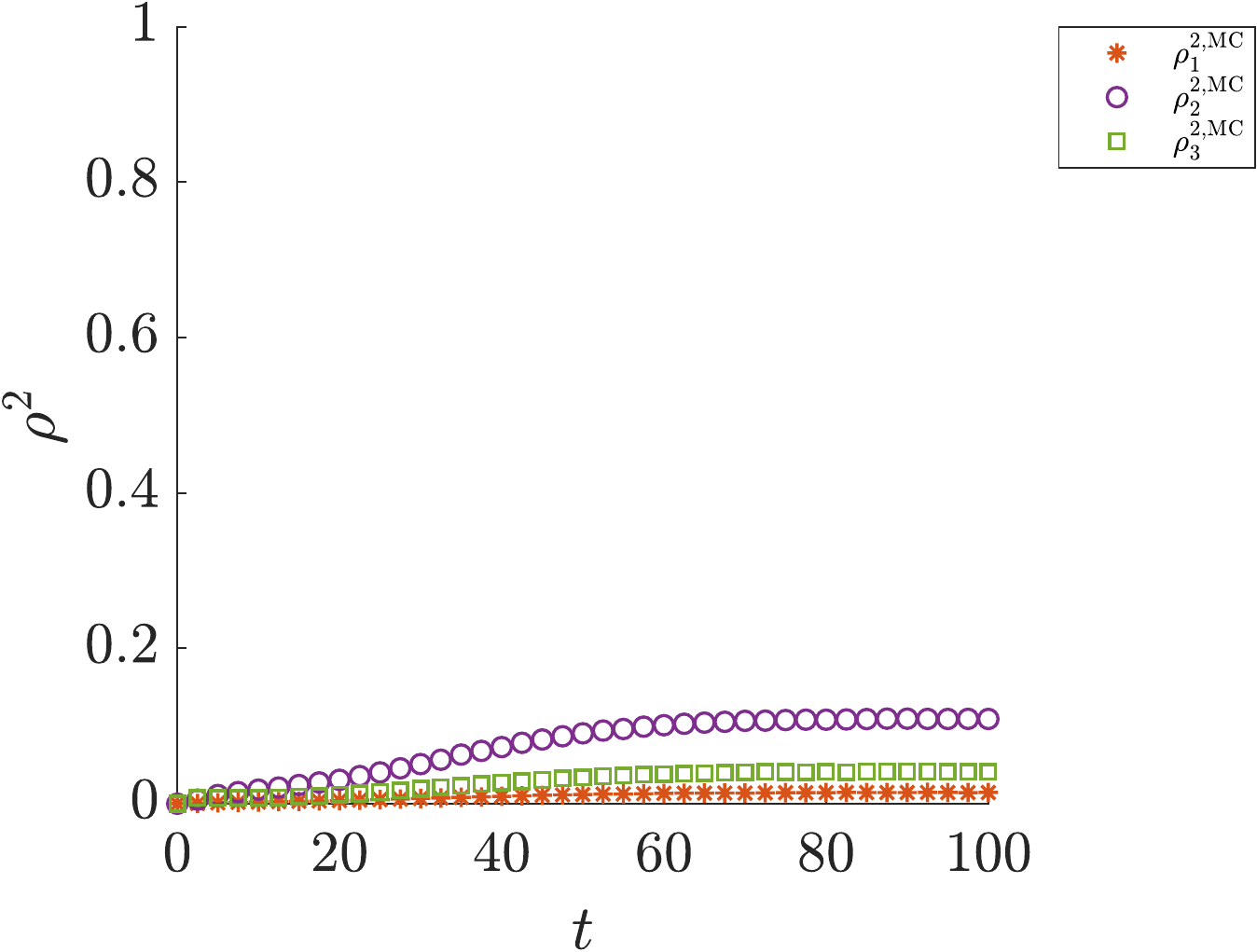}} \qquad
\subfigure[]{\includegraphics[width=.47\textwidth]{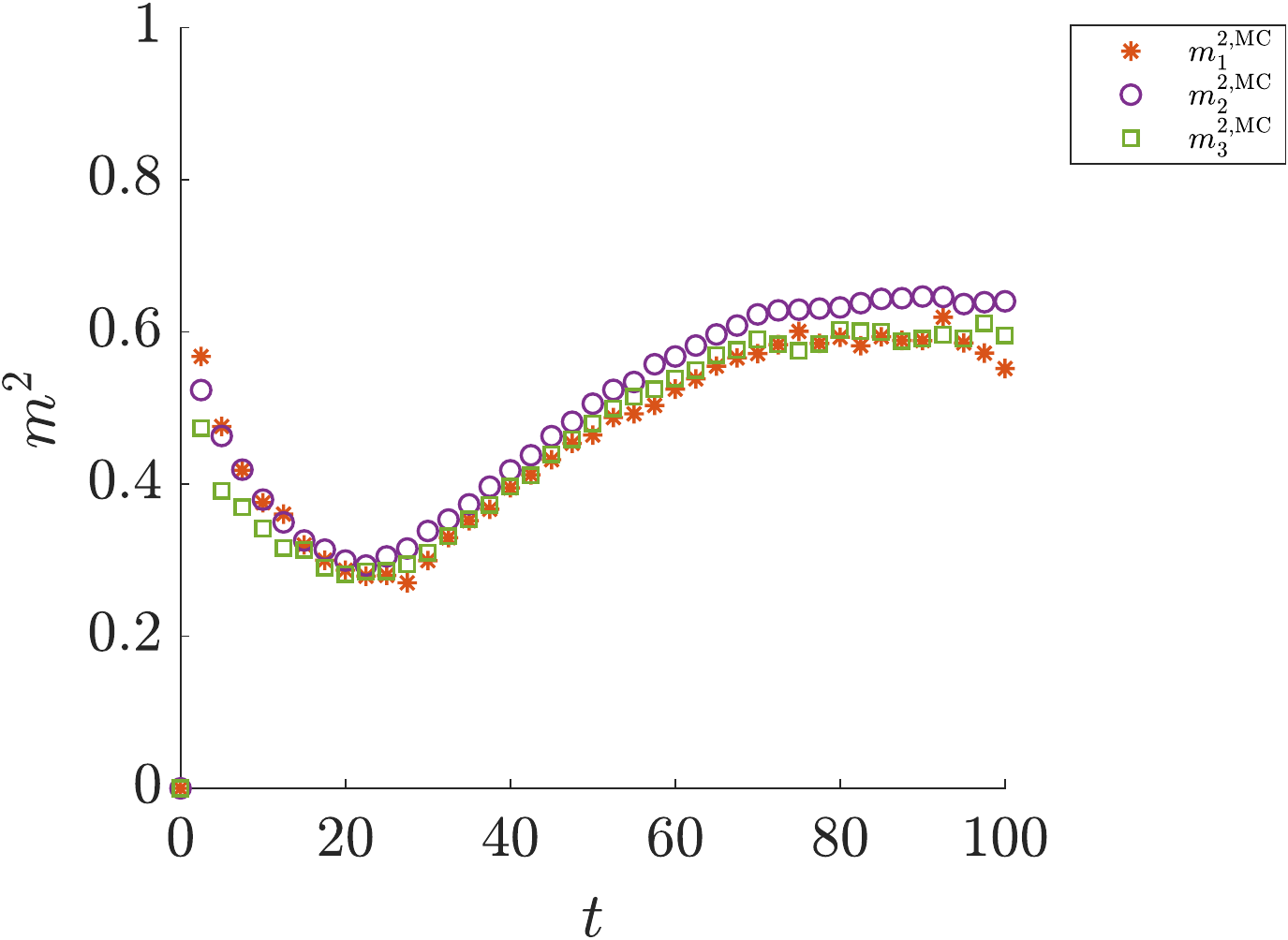}} \\
\caption{Test 3 (quarantine in the nodes, Section~\ref{sect:num.test_3}) with non-constant label switch probabilities}
\label{fig:quarantine-cp1}
\end{figure}

To conclude, we further consider the case of non-constant label switch probabilities. Specifically, we set
$$ T^{21}_i(v)=0.2(1-e^{-v}), \quad T^{12}_i(v)=0.4e^{-v}, \qquad \forall\,i\in\{1,\,2,\,3\}. $$
Hence the probability for an individual to be diagnosed as infected and quarantined increases monotonically when the viral load increases, while the probability to be recognised as recovered and readmitted in the society decreases monotonically. This is more consistent than constant label switch probabilities with the way in which real screening tests for infection detection work, see e.g.,~\cite{larremore2021SA}. Notice that these functions are bounded from above by the values formerly used as constant label switch probabilities:
\begin{equation}
	\sup_{v\in\R_+}T^{21}_i(v)=0.2, \quad \sup_{v\in\R_+}T^{12}_i(v)=0.4.
	\label{eq:sup.T}
\end{equation}
For this case we have not derived a macroscopic model in closed form (though some hints in this direction may be found in~\cite{loy2021KRM_preprint} in the networkless setting), therefore in Figure~\ref{fig:quarantine-cp1} we only show the results of the particle model~\eqref{eq:particle.quarantine}. We notice in particular that the quarantine is still successful in keeping the infection globally under control, indeed no blow-up of the mean viral load occurs in any node. Nevertheless, since the label switch probabilities are not the same for all individuals, unlike the previous case the infection cannot be eradicated by relying only on quarantine. The infection becomes instead endemic as shown by the fact that the mean viral loads reach asymptotically comparable non-zero values in all nodes, cf. Figure~\ref{fig:quarantine-cp1}b. The details of Figures~\ref{fig:quarantine-cp1}d,~\ref{fig:quarantine-cp1}f suggest furthermore that the mean viral loads of the non-quarantined individuals set reasonably on lower endemic values than those of the quarantined individuals. On the other hand, Figures~\ref{fig:quarantine-cp1}c,~\ref{fig:quarantine-cp1}e show that in the endemic regime the number of non-quarantined individuals is systematically higher than that of quarantined ones in all nodes, a fact that meets again the intuitive expectation. The lower number $\rho^2_i$ of quarantined individuals in all nodes also explains why the trend of the global mean viral loads $m_i$ (Figure~\ref{fig:quarantine-cp1}b) is quite close to that of the mean viral loads of non-quarantined individuals $m^1_i$ (Figure~\ref{fig:quarantine-cp1}d) for all $i=1,\,2,\,3$, indeed we recall that $m_i=\frac{\rho^1_i}{\rho_i}m^1_i+\frac{\rho^2_i}{\rho_i}m^2_i$.

By suitably increasing~\eqref{eq:sup.T}, i.e. the sensitivity of the testing techniques, one may expect that the model predicts the eradication of the infection in the long run also in this case.

\section{Conclusions}
\label{sect:conclusions}
In this paper, we have proposed a formal derivation of population models which describe the spreading of an infectious disease on a spatial network by taking into account the role of the viral load of the individuals. In particular, with the introduction of the viral load as a descriptive variable of the system we have modelled the transmission of the disease in a more specific way than by simply estimating the number of infectious contacts out of the gross number of susceptible and infectious individuals. Furthermore, we have avoided the necessity to partition the population of each node in infection-dependent compartments, because we have described both the individual and the aggregate epidemiological trends by means of the individual and mean viral loads, respectively.

Our derivation has taken advantage of the conceptual paradigms of statistical mechanics and kinetic theory. Starting from a particle model, in which individuals are characterised by a microscopic state comprising their viral load and their current node in the network, we have provided a mesoscopic description in terms of Boltzmann-type equations for the distributions of the viral load in the various nodes. Subsequently, thanks to the linear structure of the interaction rules modelling the pairwise transmission of viral load and the jumps across the nodes of the network, we have been able to obtain closed systems of non-linear macroscopic equations for the time evolution of the density of individuals and their mean viral load in each node. We have then characterised analytically the large-time aggregate trends, such as the existence of equilibrium density distributions and the blow-up or eradication of the infection in the nodes, in some representative cases in terms of relevant microscopic parameters. Finally, we have also extended the basic model of disease transmission on networks to even more realistic scenarios, such as the one of commuters, who move across the nodes according to more specific criteria than simply random jumps, and the one in which quarantine is applied in the nodes as a confinement measure of the infection. In all these cases we have obtained the macroscopic counterpart of the particle model via a statistical mechanics and kinetic theory approach. Our numerical tests have confirmed the matching between the particle and the macroscopic models, thereby validating the latter as reliable approximations of the former more amenable to analytical investigations and quick and accurate numerical solutions. As a by-product, our approach has provided a contribution to the broader topic of kinetic models of network-structured interactions, cf. e.g.,~\cite{burger2020PREPRINT}, by addressing Boltzmann-type kinetic equations on graphs.

Deliberately, we have not tried to match real scenarios observed during a pandemic by calibrating or comparing the results of our models with data. The present work is indeed a methodological one, specifically focused on a rigorous formal derivation of new population models, which can describe aspects normally neglected in standard epidemiological models. In this respect, our tests have essentially explored plausible prototypical scenarios, while the systematic application of our models to real case studies will be the object of future investigations. We remark however that our results already indicate the ability of our models to address interesting issues typically out of the scope of standard compartmental models. On one hand, the presence of the spatial network has reproduced the observed possible inefficiency of mobility restrictions to control the growth of epidemics, cf.~\cite{espinoza2020PLOS1}. On the other hand, the introduction of the viral load in the microscopic state of the individuals makes it possible to investigate quantitatively the interplay between sensitivity and frequency of the viral tests for an optimal screening of the population. As witnessed by the contemporary literature, see e.g.,~\cite{larremore2021SA}, this is a particularly relevant issue in the fight against newly discovered infectious disease, however still addressed in a largely qualitative and empirical way.

\section*{Acknowledgements}
This research was partially supported  by the Italian Ministry for Education, University and Research (MIUR) through the ``Dipartimenti di Eccellenza'' Programme (2018-2022), Department of Mathematical Sciences ``G. L. Lagrange'', Politecnico di Torino (CUP: E11G18000350001) and through the PRIN 2017 project (No. 2017KKJP4X) ``Innovative numerical methods for evolutionary partial differential equations and applications''.

%NL acknowledges support from INdAM (Istituto Nazionale di Alta Matematica ``F. Severi''), Italy.
NL is a postdoctoral research fellow (``titolare di Assegno di Ricerca'') of Istituto Nazionale di Alta Ma\-te\-ma\-ti\-ca (INdAM, Italy).

Both authors are members of GNFM (Gruppo Nazionale per la Fisica Matematica) of INdAM, Italy.

\bibliographystyle{plain}
\bibliography{LnTa-viral_load_network}

\begin{thebibliography}{10}

\bibitem{almeida2020PREPRINT}
L.~Almeida, P.-A. Bliman, G.~Nadin, B.~Perthame, and N.~Vauchelet.
\newblock Final size and convergence rate for an epidemic in heterogeneous
  population.
\newblock Preprint: arXiv:2010.1541, 2020.

\bibitem{apolloni2014TBMM}
A.~Apolloni, C.~Poletto, J.~J. Ramasco, P.~Jensen, and V.~Colizza.
\newblock Metapopulation epidemic models with heterogeneous mixing and travel
  behaviour.
\newblock {\em Theor. Biol. Med. Model.}, 11(3):1--26, 2014.

\bibitem{arrigoni2007CHAPTER}
F.~Arrigoni and A.~Pugliese.
\newblock Global stability of equilibria for a metapopulation {S}-{I}-{S}
  model.
\newblock In G.~Aletti, A.~Micheletti, D.~Morale, and M.~Burger, editors, {\em
  Math Everywhere}, pages 229--240. Springer, 2007.

\bibitem{barbour2004JMB}
A.~D. Barbour and A.~Pugliese.
\newblock Convergence of a structured metapopulation model to {L}evins's model.
\newblock {\em J. Math. Biol.}, 49(5):468--500, 2004.

\bibitem{bertaglia2020MMNA}
G.~Bertaglia and L.~Pareschi.
\newblock Hyperbolic models for the spread of epidemics on networks: kinetic
  description and numerical methods.
\newblock {\em ESAIM Math. Model. Numer. Anal.}, 55(2):381--407, 2020.

\bibitem{boscheri2020PREPRINT}
W.~Boscheri, G.~Dimarco, and L.~Pareschi.
\newblock Modeling and simulating the spatial spread of an epidemic through
  multiscale kinetic transport equations.
\newblock Preprint: arXiv:2012.10101, 2020.

\bibitem{burger2020PREPRINT}
M.~Burger.
\newblock Network structured kinetic models of social interactions.
\newblock Preprint: arXiv:2006.15452, 2020.

\bibitem{colizza2008JTB}
V.~Colizza and A.~Vespignani.
\newblock Epidemic modeling in metapopulation systems with heterogeneous
  coupling pattern: {T}heory and simulations.
\newblock {\em J. Theor. Biol.}, 251(3):450--467, 2008.

\bibitem{dimarco2020PRE}
G.~Dimarco, L.~Pareschi, G.~Toscani, and M.~Zanella.
\newblock Wealth distribution under the spread of infectious diseases.
\newblock {\em Phys. Rev. E}, 102(2):022303, 2020.

\bibitem{dimarco2021PREPRINT}
G.~Dimarco, B.~Perthame, G.~Toscani, and M.~Zanella.
\newblock Kinetic models for epidemic dynamics with social heterogeneity.
\newblock Preprint: arXiv:2009.01140, 2021.

\bibitem{espinoza2020PLOS1}
B.~Espinoza, C.~Castillo-Chavez, and C.~Perrings.
\newblock Mobility restrictions for the control of epidemics: {W}hen do they
  work?
\newblock {\em PLoS ONE}, 15(7):e0235731, 2020.

\bibitem{garavello2016BOOK}
M.~Garavello, K.~Han, and B.~Piccoli.
\newblock {\em Models for Vehicular Traffic on Networks}.
\newblock AIMS Series on Applied Mathematics. American Institute of
  Mathematical Sciences, Springfield, MO, 2016.

\bibitem{garavello2006BOOK}
M.~Garavello and B.~Piccoli.
\newblock {\em Traffic {F}low on {N}etworks -- {C}onservation {L}aws {M}odels}.
\newblock AIMS Series on Applied Mathematics. American Institute of
  Mathematical Sciences, Springfield, MO, 2006.

\bibitem{keeling2005JRSI}
M.~J. Keeling and K.~T.~D. Eames.
\newblock Networks and epidemic models.
\newblock {\em J. R. Soc. Interface}, 2(4):295--307, 2005.

\bibitem{kermack1991BMB}
W.~O. Kermack and A.~G. McKendrick.
\newblock Contributions to the mathematical theory of epidemics -- {I}.
\newblock {\em Bull. Math. Biol.}, 53(1--2):33--55, 1991.

\bibitem{larremore2021SA}
D.~B. Larremore, B.~Wilder, E.~Lester, S.~Shehata, J.~M. Burke, J.~A. Hay,
  M.~Tambe, M.~J. Mina, and R.~Parker.
\newblock Test sensitivity is secondary to frequency and turnaround time for
  {COVID}-19 screening.
\newblock {\em Science Advances}, 7(1):eabd5393, 2021.

\bibitem{loy2021KRM_preprint}
N.~Loy and A.~Tosin.
\newblock {B}oltzmann-type equations for multi-agent systems with label
  switching.
\newblock Preprint (doi:\texttt{10.13140/RG.2.2.11726.08001/2}), 2021.

\bibitem{martcheva2015BOOK}
M.~Martcheva.
\newblock {\em An {I}ntroduction to {M}athematical {E}pidemiology}.
\newblock Springer, 2015.

\bibitem{minc1988BOOK}
H.~Minc.
\newblock {\em Nonnegative matrices}.
\newblock Wiley-Interscience, 1988.

\bibitem{pareschi2013BOOK}
L.~Pareschi and G.~Toscani.
\newblock {\em Interacting {M}ultiagent {S}ystems: {K}inetic equations and
  {M}onte {C}arlo methods}.
\newblock Oxford University Press, 2013.

\bibitem{parino2020PREPRINT}
F.~Parino, L.~Zino, M.~Porfiri, and A.~Rizzo.
\newblock Modelling and predicting the effect of social distancing and travel
  restrictions on {COVID}-19 spreading.
\newblock Preprint: arXiv:2010.05968, 2020.

\bibitem{wesolowski2017NC}
A.~Wesolowski, E.~zu~Erbach-Schoenberg, A.~J. Tatem, C.~Lourenço, C.~Viboud,
  V.~Charu, N.~Eagle, K.~Eng{\o}-Monsen, T.~Qureshi, C.~O. Buckee, and C.~J.~E.
  Metcalf.
\newblock Multinational patterns of seasonal asymmetry in human movement
  influence infectious disease dynamics.
\newblock {\em Nat. Commun.}, 8(2069):1--9, 2017.

\bibitem{zino2021PREPRINT}
L.~Zino and M.~Cao.
\newblock Analysis, prediction, and control of epidemics: {A} survey from
  scalar to dynamic network models.
\newblock Preprint: arXiv:2013.00181, 2021.

\end{thebibliography}
\end{document}